\documentclass[%
 reprint,
 amsmath,amssymb,
 aps,
floatfix,
]{revtex4-2}

\usepackage{graphicx}
\usepackage{dcolumn}
\usepackage{bm}
\usepackage{hyperref}
\usepackage{multirow}

\begin{document}

\preprint{APS/123-QED}

\title{Small-scale dynamo with finite correlation times}

\author{Yann Carteret}
 \email{yann.carteret@gmail.com}
\affiliation{Laboratoire d’Astrophysique, EPFL, CH-1290 Sauverny, Switzerland \\}

\author{Dominik Schleicher}%
 \email{dschleicher@astro-udec.cl}
\affiliation{%
 Departamento de Astronom\'ia, Universidad de Concepci\'on, Casilla 160-C, Concepci\'on, Chile \\
}%

\author{Jennifer Schober}
 \email{schober.jen@gmail.com}
\affiliation{Laboratoire d’Astrophysique, EPFL, CH-1290 Sauverny, Switzerland}

\date{\today}

\begin{abstract}
Fluctuation dynamos occur in most turbulent plasmas in astrophysics and are the prime candidates for amplifying and maintaining cosmic magnetic fields. 
A few analytical models exist to describe 
their behaviour 
but they are based on simplifying assumptions.
For instance the well-known Kazantsev model assumes an incompressible flow that is delta-correlated in time. 
However, these assumptions can break
down
in the interstellar medium 
as it is highly compressible and the velocity field has a finite correlation time. Using
the renewing flow method developed by
Bhat and Subramanian (2014), we aim to extend Kazantsev's results to 
a more general class of turbulent flows.
The cumulative effect of both compressibility and finite correlation time over the Kazantsev spectrum is 
studied analytically. 
We derive an equation for the longitudinal two-point magnetic correlation function in real space to 
first order in the correlation time $\tau$ and for an arbitrary degree of compressibility (DOC). 
This generalised Kazantsev equation encapsulates the original Kazantsev equation. 
In the limit of small 
Strouhal numbers $St \propto \tau$ 
we use the WKB approximation to derive the growth rate and scaling of the 
magnetic power spectrum. 
We find the result that the Kazantsev spectrum is preserved, i.e. $M_k(k)\sim k^{3/2}$.
The growth rate is also negligibly affected by the finite correlation time; however, it is reduced by the finite magnetic diffusivity, and the DOC together. 
\end{abstract}

\maketitle


\section{Introduction}

The vast majority of the baryonic matter is in a plasma state, and therefore a complete description of the Universe needs to include a proper treatment of the electromagnetic force \citep{peratt1992physics}. From observations it is known that the Universe is highly magnetised. Indeed, magnetic fields are observed in almost all astrophysical bodies as for instance in asteroids \citep{hercik2020magnetic}, planets \citep{hulot2010magnetic}, stars \citep{vallee1998observations,bellot2019quiet}, galaxies \citep{krause2008magnetic,beck2012magnetic} or the intergalactic medium \citep{vikhlinin2002cold,han2017observing,van2019diffuse}. Due to the broad range of objects, the typical strength and correlation length of these magnetic fields  
are distributed over several orders of magnitude.
As an example in Milky-Way like galaxies, the observed magnetic fields are of a few tens $\mu$G in strength and correlated on kilo-parsec scales \citep{beck2004role}. 

The most popular mechanism to explain the observed 
magnetic fields is the dynamo process which converts 
the kinetic energy of the flow to magnetic energy. 
In the absence of large-scale motions, 
small-scale or fluctuation dynamos \footnote{The terms ``small-scale dynamo'' and ``fluctuation dynamo'' are used interchangeably in the literature.} amplify the initial magnetic field exponentially \citep{zeldovich1990almighty};
a process which is most 
efficient on the smallest scales of the system. 
In the kinematic stage of the dynamo, the magnetic field lines are frozen
into the plasma.
Due to the turbulent motion of the flow, the action of 
the small-scale dynamo is to randomly twist, stretch, and fold these 
lines which makes the magnetic field strength grow.
However, activating the dynamo requires an already existing seed field. 
Although unclear, it is generally assumed that these seed fields 
were generated in the early Universe 
\citep{kandus2011primordial} or through astrophysical processes such as the 
Biermann battery \citep{biermann1950ursprung}. 
\citet{schober2012magnetic} also
highlighted that the small-scale dynamo can only amplify the 
magnetic field for magnetic Reynolds numbers
$R_\mathrm{M} \sim UL/\eta$ ($U$ and $L$ are respectively
the typical velocity and length scale of the system)
larger than a few hundred. 
In the non-linear regime after saturation on the smallest scales, the peak of the magnetic energy shifts from smaller to larger scales and the magnetic energy increases following a power-law \citep{schleicher2013small}. The exact behaviour of the dynamo depends on the magnetic Prandtl number 
$P_\mathrm{M}=\nu / \eta$
and on the type of turbulence 
\citep{bovino2013turbulent,schleicher2013small,BRS2022}.

The small-scale dynamo is a key process in astrophysics.  
Indeed, the strength of the magnetic fields predicted from the early Universe is not consistent with the observed typical value of a few $\mu$G in the inter-cluster medium \citep[and references therein]{carilli2002cluster} or in high redshift galaxies \citep{Bernet2008}. 
Small-scale dynamos could then provide an explanation for the fast amplification 
of magnetic fields in the radiation-dominated phase of the early Universe \citep{wagstaff2014magnetic}, in young galaxies \citep{schober2013magnetic}, and galaxy clusters \citep{zeldovich1990almighty} as they can act on time-scales much shorter than the age of the system. 
In the context of supernova-driven turbulence, it is expected to give rise to the far-infrared-radio correlation in galaxies \citep{SchleicherBeck2013} and potentially even dwarf galaxies \citep{Schleicher2016}.
Small-scale dynamos might also be involved in the 
formation
of the first stars \citep{schleicher2010small,schober2012small, Sharda2021} 
and black holes \citep{latif2014magnetic,Latif2016, Latif2022};
and thus could also affect the epoch of reionization. 

Different approaches have been published along the years in order to model the complex behaviour of dynamos. For instance \citet{adzhemyan1988turbulent} discuss the turbulent dynamo in the framework of a quantum-field formulation of stochastic magnetohydrodynamics, where it is described as a mechanism of spontaneous symmetry breaking.
An early theoretical description of the small-scale dynamo is given by \citet{kazantsev1968enhancement}. His equation describes the time evolution of the two-point magnetic correlation function under the assumption of a Gaussian incompressible flow that is delta-correlated in time. Its derivation indicates that the magnetic power spectrum scales as $M_k(k)\sim k^{3/2}$ for $q\ll k \ll k_\eta$, where $k_\eta$ is the wavenumber above which diffusion of the magnetic field dominates. Following Kazantsev's work many authors have tried to extend this model \citep[see e.g.][]{kulsrud1992spectrum,subramanian1997dynamics,schekochihin2002model,bhat2014fluctuation}. Although some astrophysical objects host plasma that is well described by an 
incompressible flow (as for instance neutron stars \citep{weber2005strange});
Kazantsev's assumptions strongly simplify the behaviour of most 
astrophysical bodies. 
Indeed the majority of the plasma in the Universe is highly 
compressible 
as indicated by observations of compressive interstellar turbulence \citep{burlaga2015situ}. 
Moreover, in realistic flows the correlation time $\tau$ should be of the order of the smallest eddy turnover time. Thus the assumptions involved in the \citet{kazantsev1968enhancement} derivation do not allow for an accurate description of all types of fluctuation dynamos. 

In this work we aim to study the small-scale dynamo for the general case of a flow that is compressible and with finite correlations in time. \citet{zeldovich1988intermittency} pointed out that the so-called renovating flows represent a solvable analytical model to study the impact of the correlation time on small-scale dynamos. In this context, \citet{bhat2014fluctuation} developed a method to study the dynamo of incompressible flows. They found that the Kazantsev spectrum was not strongly 
affected by a finite correlation time, i.e. $M_k(k)\sim k^{3/2}$. 
However, the growth rate of the dynamo is reduced. On the other hand, \citet{schekochihin2002spectra} found that a compressible flow that is delta-correlated in time also preserves the Kazantsev spectrum where compressibility also reduces the growth rate of the dynamo. As far as we know, although there are clues that the Kazantsev spectrum should be preserved in the interstellar medium (compressible and correlated in time flow), there is no previous theoretical study that demonstrates formally that the combined actions have no effect
on the $M_k(k)\sim k^{3/2}$ spectrum. 
\citet{rogachevskii1997intermittency} used a path integral method to solve the induction equation and show that a dynamo can be activated for compressible flows that are correlated in time. Their results admit solutions consistent with the Kazantsev spectrum.

The present work assumes a simplified random flow that is compressible and correlated in time. 
We present here a generalisation of the previous work 
by \citet{bhat2014fluctuation} by including the effect of 
compressibility. The paper is organised as follows: in Sec.~\ref{sec:kazantsev_theory} we briefly review the original Kazantsev theory. In Sec.~\ref{Sec_renewingflow} we present the renewing flow method used by \citet{bhat2014fluctuation}. In Sec.~\ref{sec_kazantsev_eq_renewing} we give the derivation of the original Kazantsev equation (incompressible and delta-correlated in time flow) with the use of the renovating flow method. In Sec.~\ref{sec_own_generalisation_kazantsev} we present our generalisation of the Kazantsev equation for a compressible flow that is correlated in time
and study the WKB solutions in Sec.~\ref{sec_WKB_ssolutions}. Finally, we insert our results in the current context and draw our conclusions in Sec.~\ref{sec_discussion}.

\section{Kazantsev theory}
\label{sec:kazantsev_theory}

Dynamos in the context of an isotropic flow have been hypothesised since the fifties \citep[see e.g.][]{batchelor1950spontaneous, biermann1951cosmic}; however the first one to give a complete theoretical framework was \citet{kazantsev1968enhancement}. In his work an isotropic and homogeneous flow that is delta-correlated in time was proposed. In this section we review the basics of the derivation of the Kazantsev equation and its results, in particular we follow \citet{subramanian1997dynamics} for the formalism. 

We rewrite the velocity field as 
\begin{equation}
    \boldsymbol{u} = \langle \boldsymbol{u} \rangle + \delta \boldsymbol{u},
\end{equation}
where $\langle \boldsymbol{u} \rangle$ is the mean and $\delta \boldsymbol{u}$ the fluctuations. If we assume the fluctuations to be isotropic, homogeneous, Gaussian random with zero mean and delta-correlated in time we can set the correlation function to be
\begin{equation}
    T_{ij}(r)\delta(t_1-t_2) = \langle \delta u_i(\boldsymbol{x},t_1) \delta u_j(\boldsymbol{y},t_2) \rangle ,
\end{equation}
with $r=|\boldsymbol{x}-\boldsymbol{y}|$. Any two-point correlation function can be expressed through longitudinal and transverse components \citep{de1938statistical} as 
\begin{equation}
\label{T_incomp}
    T_{ij}(r) = \hat{r}_{ij} T_\mathrm{L}(r) + \hat{P}_{ij} T_\mathrm{N}(r),
\end{equation}
with $\hat{r}_{ij} = r_i r_j / r^2$ and $\hat{P}_{ij} = \delta_{ij}-\hat{r}_{ij}$. For a divergence-free vector field (in the case of velocity: an incompressible flow $\nabla \cdot \boldsymbol{u}=0$) we can even show that the two components are related by 
\begin{equation}
\label{T_incomp2}
    T_\mathrm{N} = T_\mathrm{L} + \frac{r}{2} \frac{\mathrm{d}}{\mathrm{d}r} T_\mathrm{L}.
\end{equation}
A similar decomposition can be performed for the magnetic field. Since $\boldsymbol{B}$ is divergence-free, the magnetic correlation function can be expressed as
\begin{eqnarray}
    M_{ij}(r) &= &  \langle \delta B_i(\boldsymbol{x}) \delta B_j(\boldsymbol{y}) \rangle, \nonumber \\
    &= &  \bigl( \hat{r}_{ij} + \hat{P}_{ij} \bigl) M_\mathrm{L} + \hat{P}_{ij} \frac{r}{2}\frac{\mathrm{d}}{\mathrm{d}r} M_\mathrm{L}.
\end{eqnarray}

The time derivative of the two-point magnetic correlation function is thus given by 
\begin{eqnarray}
    \frac{\partial M_{ij}}{\partial t} &= &  \bigg\langle \frac{\partial B_i}{\partial t} B_j\bigg\rangle +\bigg\langle \frac{\partial B_j}{\partial t} B_i\bigg\rangle - \frac{\partial \langle B_i B_j \rangle }{\partial t}.
\end{eqnarray}
Inserting this expression in the induction equation 
\begin{equation}
\label{induction}
    \frac{\partial \boldsymbol{B}}{\partial t} = \boldsymbol{\nabla} \times (\boldsymbol{u} \times \boldsymbol{B}-\eta \nabla \times \boldsymbol{B}),
\end{equation}
and using the averaged induction equation
\begin{equation}
    \frac{\partial \langle \boldsymbol{B}  \rangle}{\partial t} = \boldsymbol{\nabla} \times (\langle \boldsymbol{v} \rangle \times \langle \boldsymbol{B} \rangle - [\eta+T_\mathrm{L}(0)] \nabla \langle \boldsymbol{B} \rangle),
\end{equation}
\citet{subramanian1997dynamics} found an equation for the time evolution of the longitudinal two-point magnetic correlation function
\begin{eqnarray}
\label{true_kazantsev}
    \frac{\partial M_\mathrm{L}}{\partial t} &=&  2\kappa_\mathrm{diff} M_\mathrm{L}'' + 2 \bigg( \frac{4\kappa_\mathrm{diff}}{r} + \kappa_\mathrm{diff}' \bigg) M_\mathrm{L}' \nonumber \\
    && + \frac{4}{r^2} \bigg( T_\mathrm{N} - T_\mathrm{L} - rT_\mathrm{N}' - r T_\mathrm{L}' \bigg)M_\mathrm{L}.
\end{eqnarray}
In this expression $\kappa_\mathrm{diff} \equiv \eta + T_\mathrm{L}(0) - T_\mathrm{L}(r)$ and a prime denotes a derivative with respect to $r$.
If we further suppose that the time and spatial 
dependencies are independent, we can 
use the ansatz
\begin{equation}
    M_\mathrm{L}(r,t) = \frac{1}{r^2 \sqrt{\kappa_\mathrm{diff}}} \psi (r) e^{2\Gamma t}.
\end{equation}
This form is convenient as it highlights 
a formal similarity to quantum mechanics. 
We insert the ansatz into Eq.~(\ref{true_kazantsev}) and find
\begin{equation}
    -\kappa_\mathrm{diff} \frac{\mathrm{d}^2 \psi}{\mathrm{d}r^2}+ U(r) \psi = - \Gamma \psi.
\end{equation}
This equation has the form of a Schrödinger equation and is often referred to as the Kazantsev equation in the literature, however in this work we will refer to Eq.~(\ref{true_kazantsev}) as the Kazantsev equation instead. The function $U(r)$ is equivalent to a potential and is given by 
\begin{equation}
    U(r)\equiv \frac{\kappa_\mathrm{diff}''}{2}- \frac{(\kappa_\mathrm{diff})'}{4\kappa_\mathrm{diff}} + \frac{2\kappa_\mathrm{diff}}{r^2} + \frac{2T_\mathrm{N}'}{r} + \frac{2(T_\mathrm{L}-T_\mathrm{N})}{r^2}.
\end{equation}
Note that in the derivation of this equation we did not assume at any point that the flow is incompressible.

\citet{schekochihin2002model} studied the Kazantsev equation in Fourier space in the sub-diffusion limit such that $k_f\ll k \ll k_\eta$, with $k_f$ being the forcing scale for a single scaled flow and $k_\eta$ the Fourier conjugate of the magnetic diffusion length scale (the scale at which the magnetic diffusion is important). If incompressibility is assumed, the Kazantsev equation can be rewritten as \citep[see e.g.][]{vainshtein1982theory,kulsrud1992spectrum}
\begin{equation}
    \frac{\partial M_k}{\partial t} = \frac{\gamma}{5} \bigg( k^2 
    \frac{\partial^2 M_k}{\partial k^2} -2k\frac{\partial M_k}{\partial k} + 6M_k\bigg) -2\eta k^2 M_k,
\end{equation}
where $\gamma$ is a constant that characterises the flow and
$M_k(k,t)$ represents the magnetic power spectrum. Compared to $M_\mathrm{L}(r)$ it characterises the magnetic correlation function in Fourier space, formally we have the following relation
\begin{eqnarray}
&&  \hspace{-1cm} \langle \hat{B}_i (\boldsymbol{k},t) \hat{B}_j^{\ast} (\boldsymbol{k}',t')  \rangle =  (2\pi)^3 \hat{M}_{ij}(k,t) \delta^3(\boldsymbol{k}-\boldsymbol{k}') \delta(t-t') \nonumber \\ &=& (2\pi)^3\frac{M_k(k,t)}{4\pi k^2} \bigg(\delta_{ij}-\frac{k_ik_j}{k^2}\bigg)\delta^3(\boldsymbol{k}-\boldsymbol{k}') \delta(t-t'),
\end{eqnarray}
with $\hat{A}^\ast$ being the complex conjugate of the  
Fourier transform $\hat{A}$. 
The solution of the Fourier space Kazantsev equation is given by 
\begin{align}
    M_k(k,t) = M_0 e^{\gamma \lambda t} k^{3/2} K_\mathrm{Mc} (k/k_0), && \mathrm{Mc} = \sqrt{5(\lambda-\frac{3}{4})},
\end{align}
where $K_\mathrm{Mc}$ is the Macdonald function, $\lambda$ the normalised growth rate and $k_0 = (\gamma/10)^{1/2}$. The magnetic power spectrum thus scales mostly as $M_k(k)\sim k^{3/2}$ in the sub-diffuse limit, which we refer to as the Kazantsev spectrum. The magnetic spectrum grows exponentially in time, with a growth rate given by $3\gamma / 4$ for an incompressible flow that is delta-correlated in time. 

\section{The renewing flow method}
\label{Sec_renewingflow}

The renewing or renovating flow method was firstly proposed by \citet{steenbeck1969dynamo}. 
\citet{zeldovich1988intermittency} highlighted that it provides an alternative to the unphysical assumption of 
velocities that are delta-correlated in time but 
remains analytically 
solvable. Several authors have used the method to obtain relevant results with finite correlation times \citep[e.g.][]{kraichnan1976diffusion,dittrich1984mean,haynes2005controls,kolekar2012mean,jingade2020mean}. 
In this work we employ the operator splitting method, used by \citet{gilbert1992magnetic}
to recover the mean-field dynamo equations. 
Following the approach of \citet{bhat2014fluctuation}, in a 
non-helical flow, we impose a velocity field of the form 
\begin{equation}
    \boldsymbol{u} = \boldsymbol{a} \sin{(\boldsymbol{q} \cdot \boldsymbol{x} + \psi)}.
\end{equation}
We split the time into intervals of length $\tau$ which is the correlation time of the flow. 
In each of these $\tau$-intervals we draw randomly $\boldsymbol{a}$, $\boldsymbol{q}$ and $\psi$ such that the flow is overall isotropic, 
homogeneous, and with a zero mean\footnote{We will discuss the exact 
way to draw $\boldsymbol{a}$, $\boldsymbol{q}$ and $\psi$ in future sections (see Sec.~\ref{sec_vel_param}, Sec.~\ref{sec:1stinitia} and Sec.~\ref{sec:secondini}).}. 
Note that the flow is static only in intervals of the type $[(n-1)\tau, n\tau]$ ($n$ being an integer) and renovates for each $\tau$-interval. 

In order to apply the operator splitting method we further divide the $\tau$-intervals into two sub-intervals of duration $\tau/2$. In the first one we consider that the diffusion of the magnetic field is zero but the velocity is doubled, in the second one the velocity is now set to zero and diffusion acts as twice its original value. Using the induction equation (Eq.~\ref{induction}) we need to solve the following problem
\begin{eqnarray}
&& \hspace{-1cm} \begin{aligned}
    \frac{\partial \boldsymbol{B}}{\partial t} = \boldsymbol{\nabla} \times 2\boldsymbol{u} \times \boldsymbol{B}, && t \in  [(n-1)\tau, (n-1)\tau+\tau/2],
\end{aligned} \nonumber \\
&& \hspace{-1cm} \begin{aligned}
\frac{\partial \boldsymbol{B}}{\partial t} &= - 2\eta \boldsymbol{\nabla} \times \boldsymbol{\nabla} \times \boldsymbol{B}, &&  t \in [(n-1)\tau +\tau/2, n\tau].
\end{aligned}
\end{eqnarray}
The validity and convergence of the operator splitting method is beyond the scope of this work, we refer interested readers to \citet{holden2010splitting}. 

\smallbreak

$\textbf{First sub-interval:}$ 
we consider only the ideal induction equation. In this case, due to the magnetic flux freezing, the magnetic field is given by the standard Cauchy solution \citep[see Sec.3.3 of][]{brandenburg2005astrophysical} 

\begin{equation}
    \label{mag_j}
    B_i (\boldsymbol{x},t) = \frac{J_{ij}(\boldsymbol{x}_0)}{|J_{ij}|} B_j(\boldsymbol{x}_0,t_0),
\end{equation}
where we define $\boldsymbol{x}(\boldsymbol{x}_0, t_0)$ to be the Lagrangian position at a time $t$ of a fluid element with an initial position $\boldsymbol{x}_0$ at time $t_0$. The matrix $J_{ij}$ is given by the coordinate transformation, namely 
\begin{equation}
\label{Jij}
    J_{ij} = \frac{\partial x_i}{\partial x_{0,j}},
\end{equation}
and $|\cdot|$ denotes the determinant of the matrix. 

\smallbreak

$\textbf{Second sub-interval:}$ 
we consider only the diffusion of the magnetic field. It is straightforward to solve the equation of diffusion in Fourier space 
where we denote the Fourier 
transform of $A$ by $\hat{A}$. 
We find the solution 
\begin{equation}
    \label{mag_f}
    \hat{B}_i (\boldsymbol{k}, t) = e^{-\eta \boldsymbol{k}^2 \tau} \hat{B}_j (\boldsymbol{k}, t_1),
\end{equation}
with $t_1=t_0 + \tau/2$.
\smallbreak

\noindent We express the total magnetic field evolution in Fourier space, from Eq.~(\ref{mag_j}) and Eq.~(\ref{mag_f}), as 
\begin{equation}
    \hat{B}_i (\boldsymbol{k}, t) = e^{-\eta \boldsymbol{k}^2 \tau} \int e^{- i\boldsymbol{k}\cdot \boldsymbol{x}} \frac{J_{ij}(\boldsymbol{x}_0)}{|J_{ij}|} B_j(\boldsymbol{x}_0,t_0) ~\mathrm{d}^3 \boldsymbol{x},
\end{equation}
which describes the successive 
evolution through the two sub-intervals.  

We are now ready to give an expression for the two-point correlation function of the magnetic field in Fourier space
\begin{eqnarray}
    && \hspace{-0.75cm} \left\langle \hat{B}_i (\boldsymbol{k}, t) \hat{B}^{\ast}_h (\boldsymbol{p}, t)\right\rangle = e^{-\eta  \tau (\boldsymbol{k}^2+\boldsymbol{p}^2)}\bigg\langle \int  \frac{J_{ij}(\boldsymbol{x}_0)}{|J_{ij}|} \frac{J_{hl}(\boldsymbol{y}_0)}{|J_{hl}|} \nonumber \\
    && \times B_j(\boldsymbol{x}_0,t_0) B_l(\boldsymbol{y}_0,t_0) e^{-i(\boldsymbol{k}\cdot \boldsymbol{x}-\boldsymbol{p}\cdot \boldsymbol{y})}  ~\mathrm{d}^3 \boldsymbol{x} \mathrm{d}^3 \boldsymbol{y}\bigg\rangle,
\end{eqnarray}
where $\langle \cdot \rangle$ denotes an average over the parameter space of the velocity flow and $A^{\ast}$ is the complex conjugate. We can change the integration variables $\{\boldsymbol{x},\boldsymbol{y}\} \rightarrow \{\boldsymbol{x}_0, \boldsymbol{y}_0\}$ such that the 
determinants of the two Jacobian matrices cancel. 
We can also argue that the initial magnetic field is no longer 
correlated with the renewing flow in the next sub-interval, 
which allows us to split the averages. The final expression is then given by 
\begin{eqnarray}
    \label{corr_fourier}
   && \hspace{-0.45cm}  \left\langle \hat{B}_i (\boldsymbol{k}, t) \hat{B}^{\ast}_h (\boldsymbol{p}, t) \right\rangle = e^{-\eta  \tau (\boldsymbol{k}^2+\boldsymbol{p}^2)}  \int \left\langle B_j(\boldsymbol{x}_0,t_0) B_l(\boldsymbol{y}_0,t_0) \right\rangle \nonumber \\
   && \times \left\langle J_{ij}(\boldsymbol{x}_0) J_{hl}(\boldsymbol{y}_0) e^{-i(\boldsymbol{k}\cdot \boldsymbol{x}-\boldsymbol{p}\cdot \boldsymbol{y})}\right\rangle  ~\mathrm{d}^3 \boldsymbol{x}_0 \mathrm{d}^3 \boldsymbol{y}_0 .
\end{eqnarray}
Note that in this expression $\boldsymbol{x}$ and $\boldsymbol{y}$ are functions of the initial positions. 

As the flow is overall isotropic and homogeneous we expect that for an initial state of the magnetic field, which is also isotropic and homogeneous, these properties are conserved. Under such assumptions the two-point magnetic correlation function takes the following form
\begin{equation}
\label{exp_rewite}
   \left\langle B_i (\boldsymbol{x},t) B_j (\boldsymbol{y},t)\right\rangle= M_{ij} (r,t), 
\end{equation}
where $r = |\boldsymbol{x}-\boldsymbol{y}|$. We can further introduce a new set of integration variables $\{\boldsymbol{x}_0,\boldsymbol{y}_0\} \rightarrow \{\boldsymbol{r_0} \equiv \boldsymbol{x}_0-\boldsymbol{y}_0, \boldsymbol{y}_0\}$. We rewrite the exponential part inside the integral as
\begin{equation}
    -i\left[ \boldsymbol{k}\cdot (\boldsymbol{x}-\boldsymbol{x}_0)-\boldsymbol{p}\cdot (\boldsymbol{y}-\boldsymbol{y}_0) + \boldsymbol{k} \cdot \boldsymbol{r}_0 + (\boldsymbol{k}-\boldsymbol{p})\cdot \boldsymbol{y}_0 \right].
\end{equation}
For now we assume that the evolution tensor, that is given by
\begin{equation}
\label{Rijhlexpressiongen}
    R_{ijhl} \equiv \left\langle J_{ij}(\boldsymbol{x}_0) J_{hl}(\boldsymbol{y}_0) e^{-i\left[ \boldsymbol{k}\cdot (\boldsymbol{x}-\boldsymbol{x}_0)-\boldsymbol{p}\cdot (\boldsymbol{y}-\boldsymbol{y}_0)\right] }\right\rangle,
\end{equation}
is independent of $\boldsymbol{y}_0$; which is convenient as we can rewrite Eq.~(\ref{corr_fourier}) in the following form 
\begin{eqnarray}
\label{main_eq}
   \left\langle\hat{B}_i (\boldsymbol{k}, t) \hat{B}^{\ast}_h (\boldsymbol{p}, t)\right\rangle&=&  (2\pi)^3 \delta^3(\boldsymbol{p}-\boldsymbol{k}) e^{-2\eta  \tau \boldsymbol{p}^2} \nonumber \\ 
   &&\hspace{-1.8cm} \times \int e^{-i \boldsymbol{p}\cdot \boldsymbol{r}_0} R_{ijhl} M_{jl} (r_0,t_0) ~\mathrm{d}^3 \boldsymbol{r}_0, 
\end{eqnarray}
once the integration over $\mathrm{d}^3 \boldsymbol{y}_0$ is performed. Note that the Dirac-Delta function appears from the integration over $\boldsymbol{y}_0$ since the exponential is the only dependency on $\boldsymbol{y}_0$ and can be taken out of the flow parameters average. We assumed that $R_{ijhl}$ only depends on $\boldsymbol{r}_0$ because this form of the equation is more compact; we will show in further sections (see Sec.~\ref{sec:R_ijhl incompressible}) that this assumption is valid, at least for the cases we consider. 

\section{Kazantsev equation from the renewing flow method}
\label{sec_kazantsev_eq_renewing}
In his initial work, Kazantsev considered a flow that is delta-correlated in time and incompressible. This case is the easiest to treat with equations that are more or less tractable. We use this simplified treatment to present a detailed calculation in the framework of the renewing flow method. 
With the renewing flow method we consider the velocity field to be known; which constitutes the main difference with previous works on the topic.

\subsection{Velocity flow parameters}
\label{sec_vel_param}
The first step is to give a suitable parametrisation of $\boldsymbol{a}$, $\boldsymbol{q}$ and $\psi$ to ensure the statistical isotropy and homogeneity of the flow. We further impose an incompressible flow, which translates here to the requirement that $\boldsymbol{a}$ and $\boldsymbol{q}$ are orthogonal 
to each other. 

\begin{description}
   \item[Homogeneity] we draw $\psi$ in each $\tau$-interval from a uniform distribution in the range $[0,2\pi]$.
   \item[Isotropy] we fix the value of $q$ which is the norm of $\boldsymbol{q}$. The wavenumber $\boldsymbol{q}$ is randomly drawn from a sphere of radius $q$. The velocity orientation $\boldsymbol{a}$ is randomly drawn in the plane perpendicular to $\boldsymbol{q}$ such that $\langle \boldsymbol{u} \rangle$=0. 
\end{description}

\noindent In order to simplify the computations we change the average ensemble. Instead of averaging over the direction of $\boldsymbol{a}$ we prefer to use a new vector $\boldsymbol{A}$ which has a fixed norm and a direction drawn randomly. Then $\boldsymbol{A}$ and $\boldsymbol{q}$ define a plane where we can project the component of $\boldsymbol{A}$ that is orthogonal to $\boldsymbol{q}$. This is performed by 
\begin{align}
\label{form_a}
    \tilde{P}_{ij} \equiv \delta_{ij} - \hat{q}_i \hat{q}_j, && a_i = \tilde{P}_{ij} A_j,
\end{align}
where $\hat{q}_i \equiv q_i/q$ is the normalised component of $\boldsymbol{q}$. Note also that we adopt the Einstein summation rule. Since $\boldsymbol{A}$ and $\boldsymbol{q}$ are two independent vectors this parametrisation ensures $\langle \boldsymbol{u} \rangle= 0$. We directly see that $a$ is not fixed in this context, however we can evaluate it from $A$ as 
\begin{eqnarray}
\label{norm_a}
   \hspace{-0.3cm} \left\langle a^2 \right\rangle &=&  \left\langle a_i a_i \right\rangle =\left\langle \tilde{P}_{il} A_l \tilde{P}_{ih} A_h \right\rangle, \nonumber \\  
   \hspace{-0.3cm} &\underset{\text{average of } \boldsymbol{A}}{=} &  \frac{A^2}{3}\left\langle \tilde{P}_{il} \tilde{P}_{ih} \delta_{lh} \right\rangle \underset{\text{average of } \boldsymbol{q}}{=} \frac{2A^2}{3},
\end{eqnarray}
where we used the fact that $\langle A_i A_j \rangle= \delta_{ij}/3$ for a random vector. 

\subsection{Two-point velocity correlation functions}

In order to reconstruct the original 
Kanzatsev equation~\ref{true_kazantsev} 
we only need to compute the second-order velocity correlator. We use the definition of \citet{bhat2014fluctuation}
\begin{equation}
\label{eq_T}
    T_{ij} = \frac{\tau}{2}\left\langle u_i (\boldsymbol{x}) u_j(\boldsymbol{y})\right\rangle \underset{\text{average of } \psi}{=} \frac{\tau}{4}\left\langle a_i  a_j \cos{(\boldsymbol{q}\cdot \boldsymbol{r})}\right\rangle.
\end{equation}
The factor $\tau/2$ is required here as the flow is correlated in time. It also ensures that in the limit $\tau \rightarrow 0$ we recover the Kazantsev equation. The initialisation of $\boldsymbol{a}$ and $\boldsymbol{q}$ allows us to give an exact formula for this correlator. Using Eqs.~(\ref{form_a}) and~(\ref{norm_a}) we average over the directions of $\boldsymbol{A}$ to eliminate it, the remaining average is thus only over the directions of $\boldsymbol{q}$ with
\begin{eqnarray}
    T_{ij} &=& \frac{\tau}{4}\left\langle\tilde{P}_{il} A_l \tilde{P}_{jh} A_h \cos{(\boldsymbol{q}\cdot \boldsymbol{r})}\right\rangle= \frac{A^2\tau}{12}\left\langle\tilde{P}_{ij} \cos{(\boldsymbol{q}\cdot \boldsymbol{r})}\right\rangle \nonumber \\ &=& \frac{a^2\tau}{8} \left[ \delta_{ij} + \frac{1}{q^2} \partial_i \partial_j \right] \left\langle\cos{(\boldsymbol{q}\cdot \boldsymbol{r})}\right\rangle,
\end{eqnarray}
where we use the following notation $\partial_i \equiv \partial/ \partial r_i$. If we recall the proper definition of the average we can write 
\begin{eqnarray}
   \left\langle\cos{(\boldsymbol{q}\cdot \boldsymbol{r})}\right\rangle &\equiv &  \frac{1}{4\pi} \int^{2\pi}_0 \int^{\pi}_0 \sin{(\theta)} \cos{(\boldsymbol{q}\cdot \boldsymbol{r})}~\mathrm{d}\theta \mathrm{d} \phi \nonumber \\ &=& \frac{1}{2} \int^{\pi}_0 \sin{(\theta)} \cos{(qr \cos{(\theta)})}~\mathrm{d}\theta \nonumber \\ &=&  j_0(qr),
\end{eqnarray}
where $j_0(x)$ is the spherical Bessel function. 

\subsection{Computation of the evolution tensor}
\label{sec:R_ijhl incompressible}
In order to evaluate $R_{ijhl}$ we first need to have an expression for $J_{ij}$. As we required that $\boldsymbol{a}$ and $\boldsymbol{q}$ are orthogonal we have 
\begin{equation}
     \frac{\mathrm{d}(\boldsymbol{q} \cdot \boldsymbol{x} + \psi)}{\mathrm{d}t} \equiv \frac{\mathrm{d}\phi}{\mathrm{d}t}= 2 \boldsymbol{q}\cdot\boldsymbol{u}=0,
\end{equation}
such that $\phi$ is constant along the trajectory of a fluid element. So the equation $\mathrm{d} \boldsymbol{x}/ \mathrm{d}t=2\boldsymbol{u}$ can be easily integrated\footnote{The factor 2 comes from the fact that in the first sub-interval we have twice the initial velocity.} and gives 
\begin{equation}
    x_i - x_{0,i} = a_i \tau  \sin{(\boldsymbol{q}\cdot \boldsymbol{x}_0 + \psi)}
\end{equation}
for the Lagrangian positions. Using Eq.~(\ref{Jij}), it is straightforward to evaluate $J_{ij}$; from the last relation 
\begin{equation}
    J_{ij} (\boldsymbol{x}_0) = \delta_{ij} + \tau a_i q_j \cos{(\boldsymbol{q}\cdot \boldsymbol{x}_0 + \psi)}.
\end{equation}

\citet{bhat2014fluctuation} motivated an expansion of the exponential of the evolution tensor (Eq.~\ref{Rijhlexpressiongen}) in the limit of 
small Strouhal numbers $St = qa\tau \ll 1$.
In the context of small-scale turbulent dynamos the magnetic spectrum in the kinematic regime peaks around the resistive scale \citep{schekochihin2004simulations,iskakov2007numerical,schober2015saturation} which can be evaluated to be 
$r_{\eta} \sim (l_0/R_\mathrm{M}^{1/2})$ with $l_0$ being 
the integral scale of the flow\footnote{This definition for the resistive scale is limited to the cases $R_\mathrm{e} \simeq 1$. Some authors used a more general expression $k_\eta \propto R_\mathrm{M}^{1/2} R_\mathrm{e}^{1/4}$ \citep{schekochihin2002spectra,kriel2022fundamental,brandenburg2023dissipative}}. 
In the case considered here, the flow has only one typical scale ($1/q$), thus $r_{\eta} \sim 1/(qR_\mathrm{M}^{1/2})$. We used $R_\mathrm{M}\sim a/(q\eta)$ for the magnetic Reynolds number which is usually very high in astrophysical objects \citep[see Tab.~1 of][]{brandenburg2005astrophysical} such that $q r_{\eta}$ is very small and hence $\sin{(\boldsymbol{q}\cdot \boldsymbol{r_{\eta}})}\sim \boldsymbol{q}\cdot \boldsymbol{r_{\eta}}$. 
The phase of the exponential in Eq.~(\ref{Rijhlexpressiongen}) is then given by $aq\tau p_{\eta} r_{\eta} \sim qa\tau = St$. 
Since the terms in the vicinity of the resistive scale will contribute more to the magnetic spectrum, the expansion of $\sin{(\boldsymbol{q}\cdot \boldsymbol{x}_0 + \psi)}-\sin{(\boldsymbol{q}\cdot \boldsymbol{y}_0 + \psi)} = \sin{(\boldsymbol{q}\cdot \boldsymbol{r}_0/2)} \cos{(\boldsymbol{q}\cdot (\boldsymbol{x}_0+\boldsymbol{y}_0) + \psi)}$ is reasonable. In this section, we only keep terms up to second order in $\tau$ and we will see that it leads to the original 
Kazantsev equation~\ref{true_kazantsev}.

The equation~\ref{Rijhlexpressiongen} for $R_{ijhl}$ 
can then be rewritten in the form 
\begin{eqnarray}
\label{eq_Rinitial}
    R_{ijhl} = \bigg\langle J_{ij}(\boldsymbol{x}_0) J_{hl}(\boldsymbol{y}_0)  \big[ 1 - i\tau \beta \sigma -\frac{\tau^2 \beta^2 \sigma^2}{2!} \big]\bigg\rangle,
\end{eqnarray}
where $\beta = \sin{(\boldsymbol{q}\cdot \boldsymbol{x}_0 + \psi)}-\sin{(\boldsymbol{q}\cdot \boldsymbol{y}_0 + \psi)}$ and $\sigma = \boldsymbol{a}~\cdot~\boldsymbol{p}$. To continue further we make use of the average over $\psi$ and we also introduce the notation $\phi_{x_0}=\boldsymbol{q}\cdot \boldsymbol{x}_0 + \psi$. In fact if we try to average a function of the type 
$\cos{(n\phi_{x_0}+m\phi_{y_0})}$ or 
$\sin{(n\phi_{x_0}+m\phi_{y_0})}$ with $n$ and $m$ being two integers, 
we find that it always goes to zero except when $n=-m$. 
In particular it highlights the fact, as we hypothesised in Sec.~\ref{Sec_renewingflow}, that $R_{ijhl}$ is only dependent on $\boldsymbol{r}_0 = \boldsymbol{x}_0-\boldsymbol{y}_0$.

Term by term evaluation of the average over $\psi$ of Eq.~(\ref{eq_Rinitial}), leads
to the following expression
\begin{eqnarray}
\label{Mihrealspaceeq}
R_{ijhl} &=&  \bigg\langle  \delta_{ij}\delta_{hl} + \frac{\tau^2 a_i q_j a_h q_l }{2} \cos{(\boldsymbol{q}\cdot \boldsymbol{r_0})} \nonumber \\
&& \hspace{0.15cm} - i  \frac{\tau^2 \sigma}{2}  \sin{(\boldsymbol{q}\cdot \boldsymbol{r_0})} (\delta_{hl} a_i q_j + \delta_{ij} a_h q_l) \nonumber \\ 
&&  \hspace{0.75cm} - \frac{\tau^2 \sigma^2}{2}(1-\cos{(\boldsymbol{q}\cdot \boldsymbol{r_0})})\delta_{ij} \delta_{hl} \bigg\rangle . 
\end{eqnarray}
Each term can then be matched with Eq.~(\ref{eq_T}) to obtain
\begin{eqnarray}
    R_{ijhl} &= & \delta_{ij} \delta_{hl} - 2\tau \partial_l \partial_j T_{ih} + 2i\tau p_m (\delta_{hl} \partial_j  T_{im}  \nonumber \\
    &&  \hspace{-1.0cm} + \delta_{ij} \partial_l  T_{hm}) - 2\tau p_n p_m \delta_{ij} \delta_{hl} (T_{nm}(0)-T_{nm}),
\end{eqnarray}
where we replaced $q_i$ by suitable derivatives with respect to the components of $\boldsymbol{r}_0$ and $\sigma$ by $a_m p_m$. This expression for $R_{ijhl}$ cannot be simplified further and we need to go back to Eq.~(\ref{main_eq}) and perform the integration.

\subsection{Derivation of the Kazantsev equation}
\label{sec:Kazantsev_derivation_incompr}

The original Kazantsev equation~(\ref{true_kazantsev} 
describes the evolution of the two-point magnetic field correlation function in real space. 
Instead of evaluating $\hat{M}_{ih}(\boldsymbol{p},t)$ we take its inverse Fourier transform. Formally we get 
\begin{equation}
\label{eq_Mih_incompressible}
    M_{ih} (r,t) = \int e^{-2\eta\tau \boldsymbol{p^2}} e^{i\boldsymbol{p}(\boldsymbol{r}-\boldsymbol{r}_0)} R_{ijhl}M_{jl} (r_0,t_0)~\frac{\mathrm{d}^3\boldsymbol{r}_0 \mathrm{d}^3\boldsymbol{p}}{(2\pi)^3}.
\end{equation}

\noindent In order to further simplify this expression we assume that $\eta$ is small, such that the exponential can also be expanded giving $\exp(-2\eta \tau \boldsymbol{p}^2)\sim 1-2\eta \tau \boldsymbol{p}^2$. This expansion is justified in the context of negligible $\eta$ or large $R_\mathrm{M}$. Terms like $\eta \tau^2$ are also ignored, so the part $-2\eta \tau \boldsymbol{p}^2$ only contributes from the $\delta_{ij}\delta_{hl}$ term in the expression of $R_{ijhl}$ as it is the only term that does not depend on $\tau^2$. Once again we rewrite components of the wavevector (here $\boldsymbol{p}$) 
as derivatives
with respect to the position (here $\boldsymbol{r}$) such that $p_j \rightarrow -i \partial_j$.

We adopt the notation $[\cdot]_{ij}$ for partial derivatives with respect to $r_i$ and $r_j$. In the limit $\tau \rightarrow 0$ we can divide both sides by $\tau$ and replace $(M_{ih}(r,t)- M_{ih}(r,t_0)) / \tau \rightarrow \partial M_{ih} (r,t) / \partial t$ such that from Eq.~(\ref{eq_Mih_incompressible}) we arrive at \citep[see][for a detailed calculation]{bhat2015fluctuation}
\begin{eqnarray}
\label{time_Mih_incompressible}
     \frac{\partial M_{ih} (r,t)}{  \partial t} &=& 2\left[ M_{il} T_{jh} \right]_{jl} + 2\left[ M_{jh} T_{il} \right]_{jl}-   2\left[ M_{ih} T_{jl} \right]_{jl} \nonumber \\ 
    && \hspace{-0.28cm}- 2\left[ M_{jl} T_{ih} \right]_{jl}     +2 \left[ M_{ih} (T_\mathrm{L}(0) + \eta) \right]_{jj}.
\end{eqnarray}
Note that $T_\mathrm{L}(0)$ appears from $T_{nm} (0) = \delta_{nm} T_\mathrm{L}(0)$. This result is very important for the formalism as we started from 
equation~\ref{Mihrealspaceeq} that tracks the evolution of an initial state to the equation~\ref{time_Mih_incompressible} for the two-point magnetic field correlation function that depends on other quantities evaluated at the same space-time positions.

We can even simplify the computation further
by contracting Eq.~(\ref{time_Mih_incompressible}) with $\hat{r}_{ih}$ on both sides in order to get an equation for $M_\mathrm{L}(\boldsymbol{r},t)$. 
We would like to refer the reader to Tab.~\ref{main_table} 
for detailed expressions of different contractions that enter the computation. Using incompressibility we finally find 
\begin{eqnarray}
\frac{\partial M_\mathrm{L}(r,t)}{\partial t} &=&  \frac{2}{r^4} \partial_r \left( r^4 (\eta + T_\mathrm{L}(0) - T_\mathrm{L})\partial_r M_\mathrm{L} \right) \nonumber \\
&&- \frac{2}{r}\left( r \partial_r^2 T_\mathrm{L} + 4 \partial_r T_\mathrm{L}\right)M_\mathrm{L},
\end{eqnarray}
which is exactly the incompressible Kazantsev equation~\ref{true_kazantsev} 
in the limit of a flow that is delta-correlated in time. In comparison to previous works \citep[e.g.][]{kulsrud1992spectrum,rogachevskii1997intermittency}, the input here is the velocity field that is used to solve directly the induction equation.

\section{Generalised Kazantsev equation}
\label{sec_own_generalisation_kazantsev}
In this section we will derive the equivalent of the Kazantsev equation in the context of the renewing flow method.  
Previous studies have analysed separately the effects of the finite correlation time \citep{bhat2014fluctuation} and the compressibility \citep{schekochihin2002spectra} of the flow. By generalised we mean that we relax the incompressibility assumption used in the previous work of \citet{bhat2014fluctuation}. Our new equations then include the contributions from the time correlation of the flow as well as its degree of compressibility.

\subsection{Lagrangian positions}
\label{sec:coordtransfocompr}
In the case of an incompressible flow $\xi \equiv \boldsymbol{a} \cdot \boldsymbol{q}$ was set to 0 (see Sec.~\ref{sec_kazantsev_eq_renewing}). We can introduce a degree of compressibility by relaxing this condition; allowing $\xi$ to be non-zero with $\xi \in [-aq;aq]$. This allows us to include the non-trivial contribution from the compressibility of the flow. We can no longer apply the same reasoning as before (see Sec.~\ref{sec:R_ijhl incompressible}) since this time we have $\mathrm{d} \phi / \mathrm{d}t = 2 \xi \sin{(\phi)}$. If we integrate this expression over the first sub-interval, we find 
\begin{equation}
\label{tanphi and tanphi0}
        |\tan{(\phi/2)}| = e^{\xi \tau} |\tan{(\phi_0/2)}|.
\end{equation}
We defined $\phi = \boldsymbol{q} \cdot \boldsymbol{x} + \psi $ to be the phase of the velocity field at the final position (after a time $\tau/2$) and $\phi_0 = \boldsymbol{q} \cdot \boldsymbol{x}_0 + \psi$ to be the phase of the initial position.

Furthermore by integrating the velocity field we get
\begin{equation}
\label{coord_transfo_compr}
        x_i - x_{0,i} = \int \frac{\mathrm{d} x_i}{\mathrm{d} t}~\mathrm{d}t = \frac{a_i}{\xi} ( \phi - \phi_0).
\end{equation}
However this formula cannot be inverted. The idea is thus to use Eq.~(\ref{tanphi and tanphi0}) to isolate $\phi$ in order to plug it into Eq.~(\ref{coord_transfo_compr}) such that we get an expression of the Lagrangian positions $\boldsymbol{x}$ that depends only on the initial position $\boldsymbol{x}_0$.

We have imposed a peculiar velocity field that is periodic with respect to the variable $\phi$ with a period of $2\pi$. It is then expected that the displacement $\boldsymbol{x}-\boldsymbol{x}_0$ also possesses this periodicity. Furthermore, the velocity field is static in a $\tau$-interval which means that fluid elements are permanently pushed in the direction of $\boldsymbol{a}$ until 
they reach a zero of the velocity field 
and stop moving.
As a result a fluid element with initial position 
$\phi_0 \in [n \pi,(n+1)\pi]$ will have a position after a time $\tau/2$ such that $\phi \in [n \pi,(n+1)\pi]$ where $n$ 
is an integer. 
Eq.~(\ref{tanphi and tanphi0}) can thus be inverted, leading to 
\begin{equation}
\label{phi_inter}
    \phi/2 - \pi \lfloor \phi/(2\pi) + 1/2 \rfloor= \arctan{\left( e^{\xi \tau} \tan{(\phi_0/2)} \right)}.
\end{equation}
Recall that in Sec.~\ref{sec:R_ijhl incompressible} we motivated an expansion with respect to a small Strouhal number $St$. We motivate the same idea here as $|\xi \tau| = |aq \tau \cos{(\gamma)}| < St$ with $\gamma$ being the angle between $\boldsymbol{a}$ and $\boldsymbol{q}$. With a similar argument we can show that any new term depends directly on $St$ raised to some higher powers. In order to include effects due to finite correlation times we keep terms up to fourth order in $\tau$. The expansion of the right hand side of Eq.~(\ref{phi_inter}) also harbours a floor function that will cancel the one on the left hand side.

\smallbreak
We are now ready to plug the expression of $\phi$ from this expansion into Eq.~(\ref{coord_transfo_compr})
\begin{eqnarray}
\label{x}
     x_i & = & x_{0,i} + a_i \tau  \biggl( \sin{(\phi_0)}+ \frac{\xi \tau}{4} \sin{(2\phi_0)}  \nonumber \\
    &&+ \frac{\xi^2 \tau^2}{12} \bigl(\sin{(3\phi_0)}- \sin{(\phi_0)}\bigl) \nonumber \\
    && +   \frac{\xi^3 \tau^3}{96} \bigl(3\sin{(4\phi_0)}-4\sin{(2\phi_0)}\bigl)\biggl) ,
\end{eqnarray}
which has the desired limit for $\xi \rightarrow 0$. It is straightforward to show that the Jacobian is then given by 
\begin{eqnarray}
\label{newjac}
    J_{ij} & =& \delta_{ij} + a_i q_j\tau  \biggl( \cos{(\phi_0)} + \frac{\xi \tau}{2} \cos{(2\phi_0)} \nonumber \\
    &&+ \frac{\xi^2 \tau^2}{12} \bigl(3\cos{(3\phi_0)}- \cos{(\phi_0)}\bigl) \nonumber \\
    &&+   \frac{\xi^3 \tau^3}{24} \bigl(3\cos{(4\phi_0)}-2\cos{(2\phi_0)}\bigl)\biggl) .
\end{eqnarray}

\subsection{Fourth order velocity two-point function}
In order to include finite correlation times 
we have to consider terms up to the fourth order in $\tau$. 
The evolution tensor $R_{ijhl}$ is then given by 
\begin{equation}
\label{Rstyleeq}
        R_{ijhl} = \biggl\langle J_{ij} J_{hl} [ 1 - i\tau \beta \sigma -\frac{\tau^2 \beta^2 \sigma^2}{2!} +i\frac{\tau^3 \beta^3 \sigma^3}{3!} +\frac{\tau^4 \beta^4 \sigma^4}{4!} ] \biggl\rangle,
\end{equation}
where the Jacobian matrices are given in Eq.~(\ref{newjac}) and 
\begin{eqnarray}
    \beta  &=&  \sin{(\phi_x)} - \sin{(\phi_y)} + \frac{\tau \xi}{4} (\sin{(2\phi_x)} -\sin{(2\phi_y)}) \nonumber \\
    && \hspace{-0.5cm} + \frac{\tau^2 \xi^2}{12} (\sin{(3\phi_x)} -\sin{(3\phi_y)} - \sin{(\phi_x)} +\sin{(\phi_y)})  \\
    &&  \hspace{-1.1cm} + \frac{\tau^3 \xi^3}{96} (3\sin{(4\phi_x)} -3\sin{(4\phi_y)} - 4\sin{(2\phi_x)} +4\sin{(2\phi_y)})\nonumber. 
\end{eqnarray}

The evolution tensor now has dependencies on $a_i a_j a_h a_l$ due to the inclusion of $\tau^3$ and $\tau^4$ terms. It motivates the introduction of fourth order two-point correlators that are defined in \citet{bhat2015fluctuation} by
\begin{eqnarray}
&& T_{ijhl}^{x^2y^2}= \tau^2 \langle u_i(\boldsymbol{x}) u_j(\boldsymbol{x}) u_h(\boldsymbol{y}) u_l(\boldsymbol{y}) \rangle, \nonumber \\
&& T_{ijhl}^{x^3y}= \tau^2 \langle u_i(\boldsymbol{x}) u_j(\boldsymbol{x}) u_h(\boldsymbol{x}) u_l(\boldsymbol{y}) \rangle, \nonumber \\
&& T_{ijhl}^{x^4}= \tau^2 \langle u_i(\boldsymbol{x}) u_j(\boldsymbol{x}) u_h(\boldsymbol{x}) u_l(\boldsymbol{x}) \rangle,
\end{eqnarray}
where the factor $\tau^2$ is included due to the time correlation of the flow. We can carry out the average over $\psi$ such that the fourth order correlators are given by 
\begin{eqnarray}
\label{fourthordercorrgeneralexpression}
&& T_{ijhl}^{x^2y^2}= \frac{\tau^2}{8} \langle a_i a_j a_h a_l (\cos{(2\boldsymbol{q}\cdot \boldsymbol{r})} + 2) \rangle, \nonumber \\
        && T_{ijhl}^{x^3y}= \frac{3\tau^2}{8} \langle a_i a_j a_h a_l \cos{(\boldsymbol{q}\cdot \boldsymbol{r})}  \rangle, \nonumber \\
       && T_{ijhl}^{x^4}= \frac{3\tau^2}{8} \langle a_i a_j a_h a_l  \rangle.
\end{eqnarray}
\bigbreak
\noindent Note that $\boldsymbol{r}$ is still given by $\boldsymbol{r}=\boldsymbol{x}-\boldsymbol{y}$. 

Similarly to the second order velocity two-point correlation function we would like an expression for the fourth order correlators in the case of an isotropic, homogeneous and non-helical velocity field. Following the ideas of
\citet{de1938statistical,batchelor1953theory,landau2013fluid},
it can be shown that
\footnote{Note that $\overline{T}_{L/LN/N}$ is only a notation and does not describe a new average. } 
\begin{equation}
\label{fourthordercorrformex}
    T_{ijhl}(r)= \hat{r}_{ijhl} \overline{T}_\mathrm{L}(r) + \hat{P}_{(ij} \hat{P}_{hl)} \overline{T}_\mathrm{N}(r) + \hat{P}_{(ij} \hat{r}_{hl)} \overline{T}_\mathrm{LN}(r),
\end{equation}
where $\hat{r}_{ijhl}=r_ir_jr_hr_l/r^4$ and $\hat{P}_{ij}=\delta_{ij}-r_ir_j/r^2$. This formula has been derived and used by \citet{bhat2014fluctuation}. The bracket $(\cdot)$ operator denotes here the summation over all the different terms, formally $\hat{P}_{(ij} \hat{P}_{hl)}= \hat{P}_{ij}\hat{P}_{hl} + \hat{P}_{ih}\hat{P}_{jl} + \hat{P}_{il}\hat{P}_{jh}$ and $\hat{P}_{(ij} \hat{r}_{hl)}= \hat{P}_{ij}\hat{r}_{hl} + \hat{P}_{ih}\hat{r}_{jl} + \hat{P}_{il}\hat{r}_{jh}+\hat{P}_{hl}\hat{r}_{ij} + \hat{P}_{jl}\hat{r}_{ih} + \hat{P}_{jh}\hat{r}_{il}$. Knowing that it is straightforward to show that in the case of an incompressible flow the transverse, longitudinal and mixed terms are related by 
\begin{align}
    6\overline{T}_\mathrm{LN} = 2\overline{T}_\mathrm{L} + r\partial_r \overline{T}_\mathrm{L}, && 4\overline{T}_\mathrm{N}=4\overline{T}_\mathrm{LN} + r \partial_r \overline{T}_\mathrm{LN}.
\end{align}
These two relations will be especially useful to check if our 
generalised Kazantsev equation has the 
right form when assuming incompressibility. 

\subsection{Generalised equation}
The compressibility effects are characterised by the introduction of $\xi$ and $\xi^2$ in the evolution tensor and the Jacobian matrices. These factors are not necessarily fixed between two $\tau$-intervals. To treat them we just need to recall that $\xi = a_i q_i$, such that the methodology explained in Secs.~\ref{sec:R_ijhl incompressible}~\&~\ref{sec:Kazantsev_derivation_incompr} can still be applied. Surprisingly we find that the compressibility only affects the fourth order correlators, and Eq.~(\ref{time_Mih_incompressible}) still holds for a velocity field that is delta-correlated in time in the compressible case. The resulting equation is given by 
\begin{widetext}
\allowdisplaybreaks
\begin{eqnarray}
\label{fulleqdMihdt}
\frac{\partial M_{ih}}{\partial t} &=& 
  \left. 2[M_{jh}T_{il}]_{jl}-2[M_{ih}T_{jl}]_{jl}+2[M_{il}T_{jh}]_{jl}-2[M_{jl}T_{ih}]_{jl}+ 2[M_{ih} T_\mathrm{L}(0)]_{jj}
\right \} \text{ $\xi^0 \tau^2$ terms} \nonumber \\
 && \hspace{-0.3cm} \left. +2\left[ M_{ih} \eta \right]_{jj} \right\} \text{ term due to resistive exponential expansion} \nonumber \\
 && \hspace{-0.3cm} \left. +\tau \biggl(  [M_{jl}\widetilde{T}_{ihmn}]_{mnjl} + [M_{ih}(\widetilde{T}_{mnst}+T^{x^4}_{mnst}/12)]_{mnst}  \right. - [M_{jh}\widetilde{T}_{imns}]_{mnsj} - \left.[M_{il}\widetilde{T}_{hmns}]_{mnsl} \right \}\text{ $\xi^0 \tau^4$ terms} \nonumber \\
 && \hspace{0.6cm} \left.-[M_{jl}\partial_n\widetilde{T}_{ihmn}]_{mjl} - [M_{ih}\partial_t\widetilde{T}_{mnst}]_{mns}   + [M_{jh}\partial_s\widetilde{T}_{imns}]_{mnj} +[M_{il}\partial_s\widetilde{T}_{hmns}]_{mnl} \right \}  \text{ $\xi^1 \tau^4$ terms} \nonumber \\
 && \left. \hspace{0.51cm} \begin{aligned} &+\frac{2}{3}[M_{jl}\partial_m\partial_n\widetilde{T}_{ihmn}]_{jl} +\frac{2}{3} [M_{ih}\partial_s\partial_t\widetilde{T}_{mnst}]_{mn}   -\frac{2}{3} [M_{jh}\partial_n\partial_s\widetilde{T}_{imns}]_{mj} \\
    & -\frac{2}{3}[M_{il}\partial_n\partial_s\widetilde{T}_{hmns}]_{ml}-\frac{5}{48}[M_{jl}\partial_m\partial_n T^{x^2y^2}_{ihmn}]_{jl} -\frac{5}{48} [M_{ih}\partial_s\partial_t T^{x^2y^2}_{mnst}]_{mn}   \\
   & +\frac{5}{48} [M_{jh}\partial_n\partial_s T^{x^2y^2}_{imns}]_{mj} +\frac{5}{48}[M_{il}\partial_n\partial_s T^{x^2y^2}_{hmns}]_{ml}+ \frac{5}{48} [M_{ih}\partial_s\partial_t T^{x^2y^2}_{mnst}|_0]_{mn} \biggl), \end{aligned} \right\} \text{ $\xi^2 \tau^4$ terms}
\end{eqnarray}
\end{widetext}
where we introduced the tensor $\widetilde{T}_{ijhl}= T^{x^2y^2}_{ijhl}/4-T^{x^3y}_{ijhl}/3$. The compressibility adds the divergence of the fourth order velocity correlators, which is evaluated to be zero in the case of an incompressible flow; such that the incompressible limit ($\xi = 0$) gives back Eq.~(16) in \citet{bhat2014fluctuation}.

In order to derive the equation for $M_\mathrm{L}$ we contract Eq.~(\ref{fulleqdMihdt}) with $\hat{r}_{ih}$. By doing so we need to evaluate every term in the equation using the velocity correlation functions described in Eq.~(\ref{T_incomp}) and Eq.~(\ref{fourthordercorrformex}). Even if this computation does not involve very complicated algebra we will not detail our derivation as it is very extensive. However, we give in Tab.~\ref{main_table} the main tools and properties to perform the complete calculation. Another very useful expression to simplify the computation is 
\begin{eqnarray}
\partial_i \partial_j \bigl(\hat{r}_{ij} f(r) + \hat{P}_{ij} g(r)\bigl) &=& \partial^2_r f(r) + \frac{2}{r}\partial_r(2f(r)-g(r)) \nonumber \\
&& \hspace{0.3cm}+ \frac{2}{r^2}(f(r)-g(r)),
\end{eqnarray}
where $f$ and $g$ are two arbitrary functions of $r$. Using these properties, simplifications, and denoting $\overline{T}_{L/LN/N}$ the respective components of $\widetilde{T}_{ijhl}$ we eventually arrive at the generalised Kazantsev equation

\begin{widetext}
\allowdisplaybreaks
\begin{eqnarray}
\label{generalisedKazantsevequation}
        \frac{\partial M_\mathrm{L}}{\partial t} && = \partial_r^4 M_\mathrm{L} \bigg(\tau \bigl\{ \overline{T}_\mathrm{L} + \frac{T_\mathrm{L}^{x^3y}(0)}{12} \bigl\} \bigg) \nonumber \\
        && \hspace{0.12cm} + \partial_r^3 M_\mathrm{L} \bigg(\tau \bigl\{ 2\partial_r \overline{T}_\mathrm{L} + \frac{8}{r} \overline{T}_\mathrm{L} + \frac{2T_\mathrm{L}^{x^3y}(0)}{3r} \bigl\} \bigg)  \nonumber \\
        && \hspace{0.12cm} +  \partial_r^2 M_\mathrm{L} \bigg( -2T_\mathrm{L} + 2T_\mathrm{L}(0) + 2\eta + \tau \bigl\{ \frac{5}{3}\partial_r^2\overline{T}_\mathrm{L} + \frac{41}{3r} \partial_r \overline{T}_\mathrm{L} - \frac{8}{3r} \partial_r \overline{T}_\mathrm{LN} \nonumber \\
           && \hspace{1.65cm} + \frac{40}{3r^2} \overline{T}_\mathrm{L} - \frac{80}{3r^2}\overline{T}_\mathrm{LN} + \frac{32}{3r^2}\overline{T}_\mathrm{N} - \frac{5}{48} \partial_r^2 T^{x^2y^2}_\mathrm{L} -\frac{20}{48r}\partial_r T^{x^2y^2}_\mathrm{L} \nonumber \\
           && \hspace{1.65cm} + \frac{50}{48r} \partial_r T^{x^2y^2}_\mathrm{LN}-\frac{10}{48r^2} T^{x^2y^2}_\mathrm{L} + \frac{110}{48r^2} T^{x^2y^2}_\mathrm{LN} - \frac{80}{48r^2} T^{x^2y^2}_\mathrm{N} +\frac{2T_\mathrm{L}^{x^3y}(0)}{3r^2} + K \bigl\}\biggl)
        \nonumber \\
        &&\hspace{0.12cm} +  \partial_r M_\mathrm{L} \bigg( -2\partial_r T_\mathrm{L} -\frac{8}{r}T_\mathrm{L} + \frac{8}{r} T_\mathrm{L}(0) + \frac{8\eta}{r} + \tau \bigl\{ \frac{2}{3}\partial_r^3\overline{T}_\mathrm{L} + \frac{25}{3r} \partial_r^2 \overline{T}_\mathrm{L} - \frac{8}{3r} \partial_r^2 \overline{T}_\mathrm{LN} \nonumber\\
           && \hspace{1.65cm} + \frac{55}{3r^2}\partial_r \overline{T}_\mathrm{L} - \frac{104}{3r^2}\partial_r\overline{T}_\mathrm{LN} + \frac{32}{3r^2}\partial_r\overline{T}_\mathrm{N} +\frac{8}{3r^3}\overline{T}_\mathrm{L}-\frac{160}{3r^3}\overline{T}_\mathrm{LN}+\frac{64}{3r^3}\overline{T}_\mathrm{N} \nonumber \\
           &&\hspace{1.65cm} - \frac{5}{48} \partial_r^3 T^{x^2y^2}_\mathrm{L} -\frac{40}{48r}\partial_r^2 T^{x^2y^2}_\mathrm{L} 
            + \frac{50}{48r} \partial_r^2 T^{x^2y^2}_\mathrm{LN}-\frac{70}{48r^2} \partial_r T^{x^2y^2}_\mathrm{L} + \frac{260}{48r^2} \partial_r T^{x^2y^2}_\mathrm{LN} \nonumber \\
            &&\hspace{1.65cm} - \frac{80}{48r^2} \partial_r T^{x^2y^2}_\mathrm{N}-\frac{20}{48r^3} T^{x^2y^2}_\mathrm{L} + \frac{220}{48r^3} T^{x^2y^2}_\mathrm{LN} - \frac{160}{48r^3} T^{x^2y^2}_\mathrm{N} -\frac{2T_\mathrm{L}^{x^3y}(0)}{3r^3} + \frac{4K}{r} \bigl\} \biggl) \\
        && \hspace{0.12cm} +  M_\mathrm{L} \bigg(  -\frac{4}{r}\partial_r T_\mathrm{L} -\frac{4}{r}\partial_r T_{N} -\frac{4}{r^ 2}T_\mathrm{L} + \frac{4}{r^2} T_{N} + \tau \bigl\{ \frac{4}{3r}\partial_r^3\overline{T}_\mathrm{L} +\frac{4}{3r}\partial_r^3\overline{T}_\mathrm{LN} + \frac{20}{3r^2} \partial_r^2 \overline{T}_\mathrm{L}  \nonumber \\
           && \hspace{1.3cm} - \frac{12}{3r^2} \partial_r^2 \overline{T}_\mathrm{LN}- \frac{16}{3r^2} \partial_r^2 \overline{T}_\mathrm{N}+ \frac{8}{3r^3}\partial_r \overline{T}_\mathrm{L} - \frac{104}{3r^3}\partial_r\overline{T}_\mathrm{LN} + \frac{48}{3r^3}\partial_r\overline{T}_\mathrm{N} -\frac{8}{3r^4}\overline{T}_\mathrm{L}-\frac{56}{3r^4}\overline{T}_\mathrm{LN} \nonumber \\
           && \hspace{1.3cm} +\frac{80}{3r^4}\overline{T}_\mathrm{N}- \frac{10}{48r} \partial_r^3 T^{x^2y^2}_\mathrm{L} - \frac{10}{48r} \partial_r^3 T^{x^2y^2}_\mathrm{LN} -\frac{50}{48r^2}\partial_r^2 T^{x^2y^2}_\mathrm{L} 
            + \frac{30}{48r^2} \partial_r^2 T^{x^2y^2}_\mathrm{LN}+ \frac{40}{48r^2} \partial_r^2 T^{x^2y^2}_\mathrm{N} \nonumber \\
            &&\hspace{1.3cm} -\frac{20}{48r^3} \partial_r T^{x^2y^2}_\mathrm{L} + \frac{260}{48r^3} \partial_r T^{x^2y^2}_\mathrm{LN}- \frac{120}{48r^3} \partial_r T^{x^2y^2}_\mathrm{N}+\frac{20}{48r^4} T^{x^2y^2}_\mathrm{L} + \frac{140}{48r^4} T^{x^2y^2}_\mathrm{LN} - \frac{200}{48r^4} T^{x^2y^2}_\mathrm{N}  \bigl\} \biggl), \nonumber
\end{eqnarray}
\end{widetext}
where $K=5C(0)/48$ is a constant with $C(r)= \partial_r^2 (T^{x^2y^2}_\mathrm{L})+ 4\partial_r (T^{x^2y^2}_\mathrm{L})/r-10\partial_r (T^{x^2y^2}_\mathrm{LN})/r+2T^{x^2y^2}_\mathrm{LN}/r^2 - 22T^{x^2y^2}_\mathrm{L}/r^2+16T^{x^2y^2}_\mathrm{N}/ r^2$. This equation has the most generic form if we assume only isotropy, homogeneity and non-helicity of the velocity flow in the vicinity of small $St$. In order to solve this equation we should define the boundary conditions. The magnetic field correlation function should go to zero for infinitely large space scales. Also we would require $M_\mathrm{L}$ to be finite in $r=0$ such that the auto-correlation of the magnetic field is a local maxima. These two conditions can be summarised by 
\begin{align}
    \lim_{r\rightarrow 0} \partial_r M_\mathrm{L}(r,t) = 0, && \lim_{r\rightarrow \infty} M_\mathrm{L}(r,t) = 0.
\end{align}

Note that if we assume incompressibility in Eq.~(\ref{generalisedKazantsevequation}) we retrieve Eq.~(17) in \citet{bhat2014fluctuation}. 
Except for its length, 
the general aspect of the equation is unchanged for an arbitrary degree of compressibility (DOC). The most interesting difference arises in terms that depend on $M_\mathrm{L}$. In the incompressible case, these terms cancel perfectly but not when the DOC is non-zero. We can already get the intuition that these terms will control the time growth rate of the magnetic correlation function. 

\subsection{Small-scale limit}
\label{sec_big_5}
In this section we discuss the limit of length scales much smaller than the turbulent forcing scale (i.e.~$z~\equiv~qr~\ll~1$) of Eq.~(\ref{generalisedKazantsevequation}). The Kazantsev spectrum $M_k(k) \sim k^{3/2}$ is predicted in the range $q \ll k \ll k_{\eta}$. Since we consider large $R_\mathrm{M}$, it is sufficient to expand our generalised equation in the limit of small $z$. We introduce two different cases, which correspond to two initialisation for $\boldsymbol{a}$ and $\boldsymbol{q}$. The first case is used to give a detailed derivation. 
However, the second case is more general and gives rise to a lengthy calculation so we will 
only present the results.

\subsubsection{Two independent vectors}
\label{sec:1stinitia}
First consider the case where $\boldsymbol{a}$ and $\boldsymbol{q}$ are perfectly independent. It is straight forward to evaluate Eq.~(\ref{eq_T}) and Eq.~(\ref{fourthordercorrgeneralexpression}) knowing that $\langle a_i a_j a_h a_l \rangle = \delta_{(ij}\delta_{hl)}/15$ for a random vector
and we can directly plug the expansion of the correlators' components in Eq.~(\ref{generalisedKazantsevequation}).
These considerations simplify strongly our 
generalised Kazantsev equation 
such that it reduces to the following expression in the limit $z\ll 1$
\begin{widetext}
\begin{eqnarray}
\frac{\partial M_\mathrm{L}}{\partial t} &=& q^2 T_\mathrm{L}(0) \bigg[  \bigl( \frac{2\eta}{T_\mathrm{L}(0)} + \frac{z^2}{3} \bigl) \partial_z^2 M_\mathrm{L} + \bigl( \frac{8\eta}{zT_\mathrm{L}(0)} + 2z \bigl) \partial_z M_\mathrm{L} + \frac{8}{3} M_\mathrm{L} \bigg] \nonumber \\
        && + \frac{a^4 q^4 \tau^3}{160} \bigg[ \frac{z^4}{10} \partial_z^4 M_\mathrm{L} + \frac{8z^3}{5} \partial_z^3 M_\mathrm{L} + \frac{958z^2}{135} \partial_z^2 M_\mathrm{L} + \frac{404z}{45} \partial_z M_\mathrm{L} + \frac{32}{27} M_\mathrm{L} \bigg].
\end{eqnarray}
\end{widetext}

We can further assume that $\tilde{M}_\mathrm{L}$ is independent of space such that we use the 
ansatz 
$M_\mathrm{L}(r,t) = \tilde{M}_\mathrm{L}(z) e^{\gamma \tilde{t}}$ where $\tilde{t}= t T_\mathrm{L}(0) q^2$ and $\gamma$ is a normalised growth rate. We also set $\Bar{\tau} = \tau T_\mathrm{L}(0) q^2$ and rename $T_\mathrm{L}(0)=\eta_t$ to stick to the conventions used in \citet{bhat2014fluctuation}. After some algebra we end up with 
\begin{widetext}
\begin{eqnarray}
\label{1stcasesmallz}
0&=& \bigl( \frac{2\eta}{\eta_t} + \frac{z^2}{3} \bigl) \partial_z^2 \tilde{M}_\mathrm{L} + \bigl( \frac{8\eta}{z\eta_t} + 2z \bigl) \partial_z \tilde{M}_\mathrm{L} + \bigl( \frac{8}{3} - \gamma \bigl) \tilde{M}_\mathrm{L} \nonumber \\
        && + \frac{9\Bar{\tau}}{10} \bigg[ \frac{z^4}{10} \partial_z^4 \tilde{M}_\mathrm{L} + \frac{8z^3}{5} \partial_z^3 \tilde{M}_\mathrm{L} + \frac{958z^2}{135} \partial_z^2 \tilde{M}_\mathrm{L} + \frac{404z}{45} \partial_z \tilde{M}_\mathrm{L} + \frac{32}{27}\tilde{M}_\mathrm{L} \bigg].
\end{eqnarray}
\end{widetext}

We now focus
on the range $z_{\eta}=q r_\eta~\ll~z~\ll~1$, where $\Bar{\tau}$ terms cannot be neglected. 
Here, we
use a Landau-Lifshitz approximation \citep{landau2013classical} and consider $\Bar{\tau}$ to be a 
small
parameter. 
In order to derive approximated expressions for the high order derivatives of $\tilde{M}_\mathrm{L}$ as a function of the first and the second order derivatives, we neglect $\Bar{\tau}$ and $\sqrt{\eta/\eta_t}$ compared to $z$
\begin{eqnarray}
 && \hspace{-1.1cm} z^3 \partial_z^3 \tilde{M}_\mathrm{L}= -8z^2 \partial_z^2 \tilde{M}_\mathrm{L} + (3\gamma_0 -8) \partial_z \tilde{M}_\mathrm{L},\nonumber \\
        && \hspace{-1.1cm} z^4 \partial_z^4 \tilde{M}_\mathrm{L}  = (3\gamma_0+58)z^2 \partial_z^2 \tilde{M}_\mathrm{L} -10 (3\gamma_0 -14) \partial_z \tilde{M}_\mathrm{L},
\end{eqnarray}
where $\gamma_0$ is the growth rate for a delta-correlated in time flow. As a first approach (we will give in Sec.~\ref{sec_WKB_ssolutions} a more rigorous treatment) we neglect $\sqrt{\eta/\eta_t}$. The two expressions for high order derivatives can be plugged into Eq.~(\ref{1stcasesmallz}) to obtain
\begin{eqnarray}
0 &= &  \frac{9\Bar{\tau}}{10} \bigg[z^2 \bigl( \frac{3\gamma_0}{10}+ \frac{13}{135} \bigl) \partial_z^2 \tilde{M}_\mathrm{L} + z \bigl( \frac{9\gamma_0}{5}+ \frac{78}{135} \bigl) \partial_z \tilde{M}_\mathrm{L} \nonumber \\
 && \hspace{-0.8cm}+ \frac{32}{27}\tilde{M}_\mathrm{L} \bigg] + \frac{z^2}{3} \partial_z^2 \tilde{M}_\mathrm{L}  + 2z  \partial_z \tilde{M}_\mathrm{L} + \bigl( \frac{8}{3} - \gamma \bigl) \tilde{M}_\mathrm{L}  .
\end{eqnarray}
It is pretty obvious that this equation admits a power law solution $\tilde{M}_\mathrm{L} \sim z^{-\lambda}$. Solving for $\lambda$ we find 
\begin{equation}
\label{lambda1stcase}
    \lambda = \frac{5}{2} \pm \frac{i}{2} \bigg[ 4\frac{8+ 16\Bar{\tau}/5 -3\gamma}{1+ 81\gamma_0 \Bar{\tau} /100 + 13\Bar{\tau}/50} -25\bigg]^{1/2}.
\end{equation}
We find that the real part of $\lambda$ is $5/2$, which is exactly 
the same as \citet{bhat2014fluctuation} 
and is expected for a Kazantsev spectrum. 
\citet{gruzinov1996small} have argued that the growth rate, in the limit of $R_\mathrm{M} \rightarrow \infty$ is given by finding a value of $\lambda$ such that $\mathrm{d}\gamma / \mathrm{d} \lambda=0$. We can then plug that value into Eq.~(\ref{lambda1stcase}), $\gamma$ is thus given by 
\begin{equation}
    \gamma = \frac{7}{12}-\frac{147}{320}\Bar{\tau},
\end{equation}
where we used the self-consistent value for $\gamma_0=7/12$. Note that the complete expression for the growth rate of the dynamo is then
$\gamma_{\mathrm{tot}} = \gamma T_\mathrm{L}(0) q^2$.

\subsubsection{Arbitrary degree of compressibility}
\label{sec:secondini}

The main problem with the initialisation that we just presented is that it does not include any parameter to control the degree of compressibility (DOC). We define the DOC by 
\begin{equation}
    \sigma_c \equiv \frac{\langle (\nabla \cdot \boldsymbol{u})^2 \rangle}{\langle (\nabla \times \boldsymbol{u})^2 \rangle} = \frac{\langle a_i a_j q_i q_j \rangle}{\langle \epsilon_{ijk} \epsilon_{ihl} a_k a_l q_j q_h \rangle},
\end{equation}
where $\epsilon_{ijk}$ is the Levi-Civita symbol; the third expression is obtained after averaging over $\psi$. The DOC is then zero for an incompressible flow and goes to infinity for a fully irrotational flow. In order to derive an equation for an arbitrary DOC we set $\boldsymbol{q}$ to be a random vector with norm $q$ and $\boldsymbol{a}$ defined by 
\begin{equation}
    a_i = b (\tilde{P}_{ij} \hat{A}_j \sin{(\theta)} + \hat{q}_j \hat{A}_j \hat{q}_i \cos{(\theta)}),
\end{equation}
\bigbreak
\noindent where as before $\boldsymbol{A}$ is a random vector of norm $A$ and $\hat{A}_j = A_j / A$. The two parameters $b$ and $\theta$ (that are constant) allow to control, respectively, the norm of $\langle \boldsymbol{a}^2 \rangle$ and the DOC. In such a parametrisation the component of $\boldsymbol{A}$ along $\boldsymbol{q}$ is always rescaled by $\cos{(\theta)}$ whereas the component of $\boldsymbol{A}$ orthogonal to $\boldsymbol{q}$ in the plane described by $\boldsymbol{A}-\boldsymbol{q}$ is always recaled by $\sin{(\theta)}$. This parametrisation is taken for convenience, and $\theta$ can be interpreted as the mean absolute angle between $\boldsymbol{a}$ and $\boldsymbol{q}$. Although this parametrisation might seem arbitrary, we can show that the results we derive here are independent on the exact evaluation of $a_i$ as long as $\sigma_c$ is uniquely defined (see Appendix~\ref{appendix_proof_cstinde}). Under such considerations 
\begin{align}
     \sigma_c = \frac{1}{2 \tan{(\theta)}^2}, && \langle \boldsymbol{a}^2 \rangle = b^2 \bigg( \frac{2}{3} \sin{(\theta)}^2 + \frac{1}{3} \cos{(\theta)}^2 \bigg).
\end{align}
We directly see that the value $\theta=\pi/2$ represents the incompressible case and $\theta=0$ the fully irrotational one. Note that due to the random behavior of $\boldsymbol{A}$ we have $\langle \xi \rangle = 0$.

We apply the exact same methodology as for the first initialisation, which means we expand the two-point correlators, 
use the ansatz, 
and re-express the high order derivatives with the first and the second order ones. The expressions for the velocity correlators with this parametrisation can be found in the Appendix \ref{sec:appendix_initialisations} (Eq.~\ref{appendix:velocity_correlators}). Furthermore, we define the two functions 
\begin{eqnarray}
\label{eq:eps_and_zeta}
&&\hspace{-0.3cm} \epsilon(\theta) = \frac{216}{5(\Omega+3)^2} \biggl[\frac{\Omega+3}{5-\Omega}\bigg( \frac{1}{24}\Omega_1 - \frac{3}{28} \Omega_2 + \frac{3}{40}\Omega_3 \bigg) \nonumber \\ 
&& \hspace{0.5cm}\times \bigg( 5\gamma_0 -\frac{40}{\Omega+3} \bigg) + \frac{157}{420}\Omega_1 - \frac{599}{630} \Omega_2 + \frac{121}{180}\Omega_3\biggl] ,\nonumber \\
&& \hspace{-0.3cm} \zeta(\theta) = \frac{216}{5(\Omega+3)^2} \biggl[\frac{2}{3}\Omega_1 - \frac{14}{9} \Omega_2 + \frac{8}{9}\Omega_3\biggl] .
\end{eqnarray}
The set of five parameters $\Omega_{\mathrm{i}}$ appears very naturally in the derivation of Eq.~(\ref{appendix:velocity_correlators}) and depends only on $\theta$; exact expressions are given in Eq.~(\ref{appendix:param_set}). From the velocity correlators we have $\eta_t = \tau b^2 (\Omega+3)/72$. However the parameter $b$ is still free and we can set it to $b^2=6a^2 /(\Omega + 3)$ such that $\langle \boldsymbol{a}^2 \rangle = a^2$. As a result $\eta_t = \tau a^2 /12$, and the normalised correlation time can be evaluated to $\Bar{\tau} = St^2 /12$. The normalised correlation time is then fully controlled by $St$; independently on the choice of DOC.

The resulting equation is 
\begin{eqnarray}
\label{2ndcasesmallz}
 \left( \frac{2\eta}{\eta_t} + z^2\frac{5-\Omega}{5(\Omega+3)} \right) \partial_z^2 \tilde{M}_\mathrm{L}  && \nonumber \\ 
 + \left( \frac{8\eta}{z\eta_t} + 6z\frac{5-\Omega}{5(\Omega+3)} \right) \partial_z \tilde{M}_\mathrm{L}  + \left( \frac{8}{\Omega+3} - \gamma \right) \tilde{M}_\mathrm{L}&& \nonumber \\ 
  + \Bar{\tau} \bigg[ \epsilon(\theta) z^2\partial_z^2 \tilde{M}_\mathrm{L} + 6\epsilon(\theta) z \partial_z \tilde{M}_\mathrm{L} + \zeta(\theta)\tilde{M}_\mathrm{L} \bigg] &=& 0,\nonumber \\ 
\end{eqnarray}
Similarly to the first case we compute the growth rate and scale factor of the power law solution in the limit of large $R_\mathrm{M}$, we find that the real part of $\lambda$ is still $5/2$. The growth rate is given this time by 
\begin{align}
    \gamma_0 = \frac{7+5\Omega}{4(\Omega+3)}, && \gamma= \gamma_0 + \Bar{\tau} \bigl[ \zeta(\theta)- \frac{25}{4}\epsilon(\theta) \bigl].
\end{align}

We already see from this quick evaluation that the Kazantsev spectrum seems to be preserved even with an arbitrary DOC and correlation time. However to confirm this first approach and include the effects of finite magnetic Reynolds numbers, we need to study more carefully the solutions to Eq.~(\ref{2ndcasesmallz}).

\section{Finite magnetic resistivity solutions}
\label{sec_WKB_ssolutions}
The scaling solution that has been derived in the previous section only works if the term $\sqrt{\eta/\eta_t}$ is neglected. However to include effects due to a finite magnetic resistivity we should not systematically neglect it. A WKB approximation can be used to evaluate the solution of Eq.~(\ref{2ndcasesmallz}) including the finite resistivity. An explicit derivation of the WKB solutions can be found in Appendix~\ref{sec_appendix_wkb_sol}. We only review the main results obtained for the magnetic power spectrum and the growth rate of the dynamo including a finite magnetic Reynolds number. 

\begin{table}
    \centering
    \begin{tabular}{c|ccc}
    Parameters & Incompressible & Intermediate & Irrotational \\ \hline \\ [-0.85em]
    $\theta$ & $\frac{\pi}{2}$ & $\frac{\pi}{4}$ & 0 \\ [0.15em]
    $\sigma_c$ & $0$ & $\frac{1}{2}$ & $\infty$ \\  [0.25em] \hline \\ [-0.85em]
    $\lambda_k$ & $\frac{3}{2}$ & $\frac{3}{2}$ & $\frac{3}{2}$ \\  [0.25em] \hline \\ [-0.85em]
    $\gamma_0$ & $\frac{3}{4}$ & $\frac{7}{12}$ & $\frac{1}{4}$ \\ [0.25em]
    $\gamma_1$ & $-\frac{135}{224}$  & $-\frac{147}{320}$ &$-\frac{1017}{2240}$  \\ [0.25em]
    $\gamma_{R_\mathrm{M}}$ & $\frac{1}{5}+\Bar{\tau}\frac{27}{280}$  & $\frac{1}{3}+\Bar{\tau}\frac{293}{1200}$ & $\frac{3}{5}+\Bar{\tau}\frac{3429}{2800}$ \\  [0.25em] \hline \\ [-0.85em]
    $R_{\mathrm{M},\mathrm{thresh}}$ & $\sim 3 \cdot 10^5$ & $\sim 1.5 \cdot 10^5$ & $\sim 7 \cdot 10^4$ \\ [0.15em]
    \end{tabular}
    \caption{Presentation of the velocity field and magnetic spectrum parameters for three types of flow: 
    incompressible, irrotational, and intermediate (see Sec.~\ref{sec_big_5}).}
    \label{sumupparam}
\end{table}

\subsection{Growth rate}

\begin{figure}
	\includegraphics[width=\columnwidth]{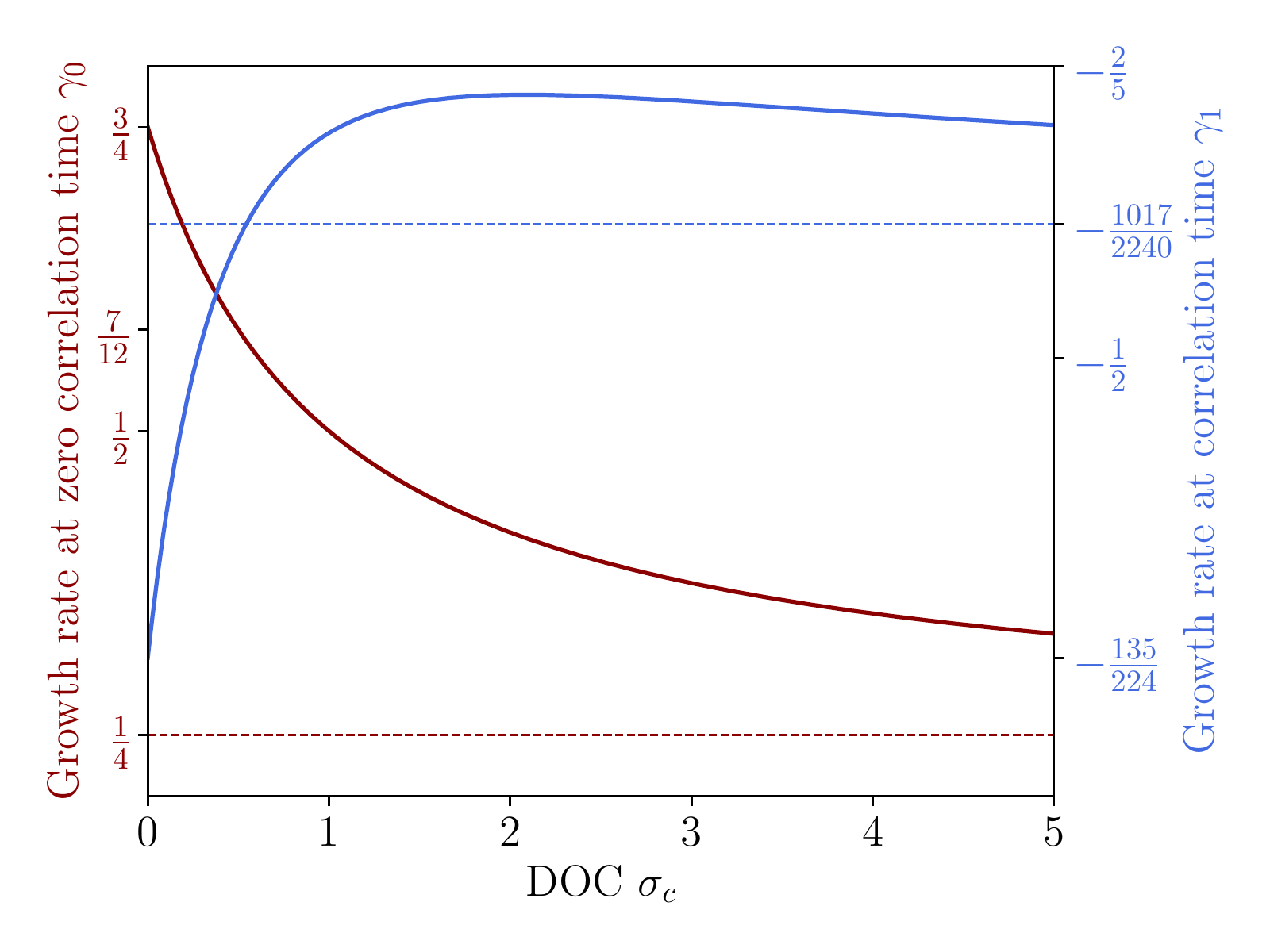}
    \caption{Evolution of the two main contributions to the magnetic spectrum growth rate with respect to the DOC for large magnetic Reynolds number. The left axis represents the main contribution $\gamma_0$ and the right axis the contribution to the growth rate related to the correlation time $\gamma_1$ of Eq.~(\ref{gamma_final_true}).
    Dashed lines correspond to the value 
    for a fully irrotational flow ($\sigma_c \rightarrow \infty$). }
    \label{fig_growthrateratio2}
\end{figure}

The normalised growth rate of the dynamo that includes contributions from the magnetic resistivity (through $R_\mathrm{M}$), compressibility (through $\Omega$ and $\theta$) and finite correlation time (through $\Bar{\tau}$) is found to be 
\begin{eqnarray}
\label{gamma_final_true}
\gamma &=& -\bigg( \frac{\pi}{ \ln{(R_\mathrm{M})}}\bigg)^2 \bigg[ \frac{5-\Omega}{5(\Omega+3)} + \Bar{\tau}\epsilon(\theta) \bigg] + \frac{7+5\Omega}{4(\Omega+3)} \nonumber\\
    && + \Bar{\tau} \bigg[ \zeta(\theta)- \frac{25}{4}\epsilon(\theta) \bigg], \nonumber\\
    &&\hspace{-0.3cm} \equiv -\bigg( \frac{\pi}{ \ln{(R_\mathrm{M})}}\bigg)^2\gamma_{R_\mathrm{M}} + \gamma_0 + \Bar{\tau} \gamma_1,
\end{eqnarray}
where the functions $\epsilon(\theta)$ and $\zeta(\theta)$ 
are given in Eq.~(\ref{eq:eps_and_zeta}) and we have 
introduced the different components of the growth rate $\gamma_{R_\mathrm{M}}$,  $\gamma_0$, and  $\gamma_1$. 
In Tab.~\ref{sumupparam} we list the different parameters of the flow and the magnetic spectrum for three regimes, namely incompressible ($\nabla\cdot \boldsymbol{u}=0$), 
irrotational ($\nabla\times \boldsymbol{u}=0$), and the 
intermediate case treated in Sec.~\ref{sec:1stinitia}. 
To get a better intuition on the results presented here, we display 
in Fig.~\ref{fig_growthrateratio2} the DOC dependency of the two 
main contributions to the growth rate. 
From the evolution of $\gamma_0$ it is very clear that 
the compressibility tends to decrease the growth rate of 
the magnetic energy spectrum of the dynamo. 
Moreover, $\gamma_0$ is comprised between 0.75 and 0.25 which indicates that the dynamo action always exists. It is interesting to note that $\gamma_0$ is a monotonously decreasing function of the DOC, whereas $\gamma_1$ has a maximum around $\sigma_c\sim 2$. 

\begin{figure}
	\includegraphics[width=\columnwidth]{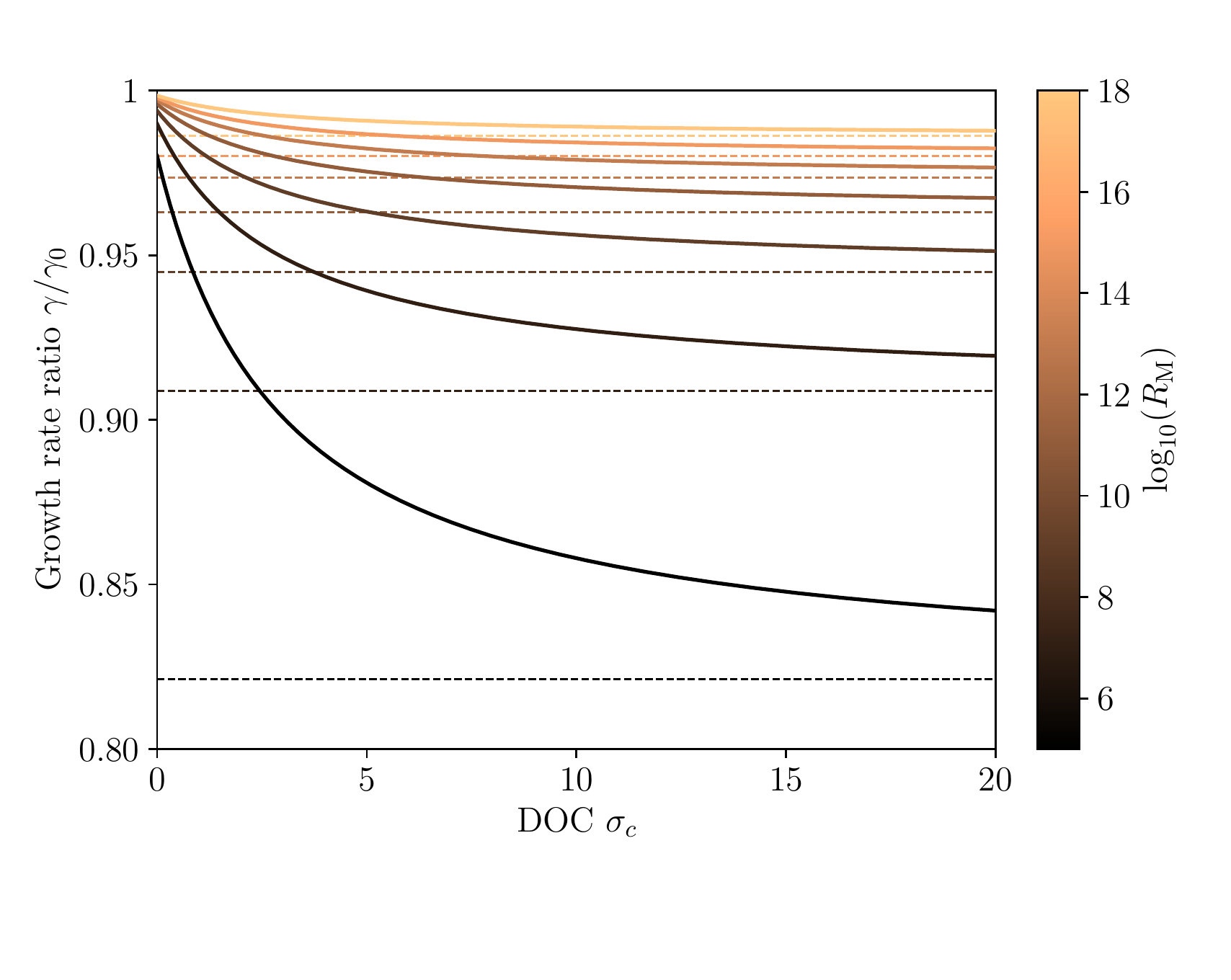}
    \caption{Ratio of the growth rate and its main contribution for a 
    few values of $R_\mathrm{M}$ and $St=10^{-2}$. 
    Dashed lines correspond to the value for a fully irrotational flow 
    ($\sigma_c \rightarrow \infty$). }
    \label{fig_growthrateratio1}
\end{figure}

In Fig.~\ref{fig_growthrateratio1} we study more precisely the dependence of the growth rate on $R_\mathrm{M}$. As expected from Eq.~(\ref{gamma_final_true}),
$\gamma$ increases as $R_\mathrm{M}$ increases.
In practice, due to the WKB approximation, there is a 
limiting value on the magnetic Reynolds number 
($R_{\mathrm{M},\mathrm{thresh}}$) for which Eq.~(\ref{gamma_final_true}) 
is valid such that we need to keep 
$R_\mathrm{M}>R_{\mathrm{M},\mathrm{thresh}}$. 
The impact on the total growth rate of the DOC is stronger when $R_\mathrm{M}$ is small. 
In this work, we only considered first order corrections; 
and discrepancies can already represent $\sim~85\%$ for 
the lowest values of $R_\mathrm{M}$ presented. 
However, the correlation time has a negligible impact 
on the total growth rate in the limit $St \ll 1$. 
Indeed, the correlation time enters the computation 
through $St$ which itself contributes through 
$\Bar{\tau} \propto St^2 \ll 1$. 

\subsection{Magnetic power spectrum}
In the range of interest, $z_{\eta}~\ll~z~\ll~1$, we find the solution for the longitudinal two-point magnetic correlation function

\begin{equation}
\label{MLwetestlplp}
    M_\mathrm{L}(z,t) = e^{\gamma \tilde{t}} z^{-5/2} M_0 \cos{\bigg[ \frac{\pi}{\ln{(R_\mathrm{M})}} \ln{\bigg( \frac{z}{z_0} \bigg)} \bigg]},
\end{equation}
where $\gamma$ is given by Eq.~(\ref{gamma_final_true}). This result is already very interesting as we can identify a power law $z^{-5/2}$ independent of the correlation time that dominates the spectrum compared to the slowly varying $\cos{[\ln{(z)}]}$ function. However the Kazantsev spectrum we are interested in predicts that the magnetic power spectrum scales as $M_k(k)\sim k^{-3/2}$ in the range $q\ll k\ll k_\eta$. We can show (see Appendix~\ref{appforMltoM}) that the magnetic power spectrum and the longitudinal two-point correlation function are related by 
\begin{equation}
    \label{forMltoM}
    M_k(k,t) = \frac{1}{\pi} \int (kr)^3 M_\mathrm{L}(r,t) j_1(kr)~\mathrm{d}r.
\end{equation}
The Bessel function $j_1(x)$ is very peaked around $x\sim 2$, so the dominant part of the integral is around $k\sim 1/r$. Thus for $M_\mathrm{L}(r)\sim r^{\lambda}$ we have $M_k(k)\sim k^{\lambda_k}$ with $\lambda_k = -(1+\lambda)$. Plugging $\lambda = -5/2$ gives the well-known Kazantsev spectrum $M_k(k)\sim k^{3/2}$ even for the compressible and time-correlated flow considered here. Note that the main contribution to the power spectrum derived does not depend on the any of the parameters of the flow. The last row of Tab.~\ref{sumupparam} corresponds to the minimal value of $R_\mathrm{M}$ for which the WKB approximation holds (see Appendix~\ref{sec:validity}). We see that for most astrophysical application of the small-scale dynamo our derived results remain valid.

\section{Discussion and conclusions}
\label{sec_discussion}

Several authors have previously modeled the kinematic phase of the small-scale 
dynamo with the Kazantsev theory
\citep[see e.g.][]{kulsrud1992spectrum,schekochihin2002spectra,schekochihin2002model,martins2019kazantsev}. 
They found that the Kazantsev spectrum is preserved, even for a compressible flow.  
However, they often assumed Gaussian statistics of the velocity field, 
such that the flow is delta-correlated in time.
There are several examples of analytic treatments that include a finite correlation time:  
\citet{kolekar2012mean} and \citet{lamburt2001fast} used a similar approach to our work but 
in the context of the mean-field dynamos, 
\citet{schekochihin2001finite} and \citet{kleeorin2002magnetic} 
considered a general case of fluctuation dynamo, and \citet{bhat2015fluctuation} solved the incompressible case.
Most of the theoretical studies, if not all, 
have found that the Kazantsev spectrum is 
preserved even if compressibility, finite correlation time, or finite resistivity are considered. 
Our work shows that the combined effect of the three 
on the Kazantsev spectrum is negligible, 
as Eq.~\ref{MLwetestlplp} scales mostly with the power-law $z^{-5/2}$. 
However, our results are only derived for first order 
corrections from the correlation time; 
as higher order corrections are usually hard to treat.

Besides the shape of the magnetic energy spectrum, the dynamo growth rate $\gamma$ is of particular interest. 
Our results for $\gamma$ are similar to the ones obtained by \citet{kulsrud1992spectrum} and \citet{schekochihin2002model} 
for the limit of incompressibility. 
\citet{schekochihin2002spectra} also derived a formula for the growth rate for an arbitrary DOC for a delta-correlated in time flow.
They found that $\gamma$ ranges between $3/4$ for an incompressible flow and $1/8$ for a fully irrotational flow. It does not match our results by a factor of two in the limit of the fully irrotational flow. Similarly, the growth rate related to the finite resistivity $\gamma_{R_\mathrm{M}}$ matches their result in the incompressible case but is overestimated by a factor of 2 in the fully compressible limit.
The discrepancy can be solved when we consider instead of $\gamma$ alone the complete growth rate, namely $\gamma \eta_t q^2$. 
In their paper \citet{schekochihin2002spectra} defined the initial growth rate from the velocity correlators in Fourier space while we define it from real space. 
If we transfer a factor $(\Omega+3)/4$ from $\eta_t$ to $\gamma$, our growth rate matches theirs in both limits. It is worth mentioning that  \citet{illarionov2021finite} derived a growth rate that depends on the Strouhal number but does not match ours exactly. However, they also found that $\gamma_{\mathrm{tot}} \propto aqSt$ and that $\bar{\tau}$ reduces the growth rate.
Although, we derived a expression for the growth rate 
that includes the DOC, a finite resistivity and time correlations our treatment of turbulence is very simple due to the imposed velocity field. A more rigorous treatment \citep{schober2012magnetic} highlights that the growth rate might also be a power law of the Reynolds number. 
\citet{rogachevskii1997intermittency} used a different approach to the problem as they impose directly a velocity correlation function instead of the velocity field itself. Their results match ours for the magnetic power spectrum as long as $T_\mathrm{L}(r) \sim r^2$; but the growth rate differs as their velocity spectrum is different from ours. 
This highlights 
that the Kazantsev spectrum should be preserved for a large class of flows. 
Using a similar approach, some authors \citep{vergassola1996anomalous,vincenzi2002kraichnan,arponen2007dynamo} studied more deeply the influence of the Prandtl number on the growth rate or magnetic spectrum while we fix the Reynolds number in our work. These papers suggest that the growth rate we derived should also depend on the Prandtl number if an even more general case is considered; this relation is beyond the scope of this paper as we want to focus on the effect of compressibility.

From a numerical point of view it seems indeed that 
a slope close to $3/2$ in the magnetic energy spectrum 
can be observed in both incompressible and compressible 
MHD simulations at large length scales \citep[see e.g.][]{haugen2004simulations,federrath2014turbulent,federrath2016magnetic,brandenburg2022batchelor,kriel2022fundamental}. 
Although the slope measured in simulations is often close 
to the theoretical prediction, 
small discrepancies can still arise. 
One of the possible explanations can directly appear 
from the size of the simulation box. 
Indeed $k$ has to be small but simulations are limited in resolution. 
This often can lead to an insufficient separation of spatial scales. 
A related problem is the assumption of very large 
hydrodynamic and/or magnetic Reynolds numbers in theoretical models. 
The required large values of these two numbers make 
a comparison between numerical simulations and theory hard.
\citet{kopyev2022magnetic} also found that time irreversible flows can generate a 
nontrivial 
deviation to the Kazantsev spectrum.
Regarding the growth rate of the dynamo, its reduction by the correlation time has also been observed in numerical studies \citep{mason2011magnetic}. 
Further discussions of the current state of dynamo numerical simulations can be found in \citet{brandenburg2012current}. 

In conclusion, we have given an example of an analytical treatment 
for the fluctuation dynamo in the most generic case of a compressible flow with a finite correlation time. 
To this end, we proposed a framework to study the cumulative 
effects of a finite correlation time and an arbitrary degree of compressibility by generalising the former work of \citet{bhat2014fluctuation}. We used the renovating flow method which assumes a very crude flow that does not allow for a very complex modelling of turbulence but keep the analytical treatment tractable. We derived a generalisation to the Kazantsev equation in real space (Eq.~\ref{generalisedKazantsevequation}) that is valid at any scale. We note however that if we assume an incompressible flow that is delta-correlated in time at this point we retrieve the original Kazantsev equation. This equation describes the time evolution of the two-point magnetic correlation function $M_\mathrm{L}$ from the velocity correlators and the spatial derivatives of $M_\mathrm{L}$ up to the fourth order. We then studied solutions for length scales much smaller than turbulent forcing scale (i.e.~$qr \ll 1$).

By the use of the WKB approximation,
we derived formulas for the growth rate and 
slope of the magnetic power spectrum $M_k(k)$ 
for large magnetic Reynolds number $R_\mathrm{M} \gg 1$ and small Strouhal number $St \ll 1$. 
In particular, it allowed to capture the effect of finite magnetic diffusivity. Furthermore we could define a lower bound on $R_\mathrm{M}$ for which our results should hold, $R_{\mathrm{M},\mathrm{thresh}}\sim 10^5$, which is smaller than most of the typical values in astrophysical objects. Although the growth rate showed dependencies on both the degree of compressibility and the correlation time, the Kazantsev spectrum seemed to be preserved, 
i.e.~$M_k(k)\sim k^{3/2}$, independently of $\tau$ or $\sigma_c$. 
Our results are derived in a very special context, namely for a renovating flow.
But our predictions regarding the magnetic field spectrum seem robust in the 
sense that both numerical and theoretical studies agree with the conservation 
of the Kazantsev spectrum for compressible and time-correlated flows.

\begin{acknowledgments}
DRGS gratefully acknowledges support by the ANID BASAL projects ACE210002 and FB210003, as well as via the Millenium Nucleus NCN19-058 (TITANs). YC and DRGS thank for funding via Fondecyt Regular (project code 1201280). 
JS~acknowledges the support by the Swiss National Science Foundation under Grant No.\ 185863.
\end{acknowledgments}

\appendix

\section{General initialisation for delta-correlated in time flow}
\label{appendix_proof_cstinde}
We would like to review an even more general initialisation than the one presented in Sec.~\ref{sec:secondini}. We only consider a delta-correlated in time flow but this discussion could in principle be generalised to a finite correlation time. A very general expression for $\boldsymbol{a}$ that preserves isotropy is
\begin{equation}
    a_i = b(\tilde{P}_{ij}\hat{A}_j f_1+ \hat{q}_j \hat{A}_j \hat{q}_i f_2),
\end{equation}
where $f_1$ and $f_2$ are two constants. In order to control the norm $a$ we should impose that $f_{1/2}$ are between minus one and one. It is then straightforward to show that $\sigma_c= f_2^2 /(2f_1^2)$. Once again we compute the velocity correlators and plug in Eq.~(\ref{generalisedKazantsevequation}). To simplify this derivation we also neglect the resistivity $\eta$. We find the equation
\begin{eqnarray}
&& \frac{z^2}{5}\frac{2f_1^2+3f_2^2}{2f_1^2+f_2^2} \partial_z^2 \tilde{M}_\mathrm{L} + \frac{6z}{5}\frac{2f_1^2+3f_2^2}{2f_1^2+f_2^2} \partial_z \tilde{M}_\mathrm{L}\nonumber \\
&&\hspace{1.5cm}+ \big( 4\frac{f_1^2+f_2^2}{2f_1^2+f_2^2} -\gamma \big) \tilde{M}_\mathrm{L} = 0,
\end{eqnarray}
that again allows some power law solution. If we follow the same approach than in Sec.~\ref{sec:secondini}, we find 
\begin{align}
    \lambda = \frac{5}{2} \pm i g(f_1,f_2,\gamma), && \gamma = \frac{6f_1^2+f_2^2}{4(2f_1^2+f_2^2)},
\end{align}
with $g(f_1,f_2,\gamma)$ a function that characterises the growth rate. Once again the power spectrum slope is constant and $\gamma = 3/4$ for an incompressible flow, $\gamma=1/4$ for a fully irrotational one and $\gamma = 7/12$ if $f_1=f_2$. If we define 
\begin{align}
    f_1 = \sin{\theta}, && f_2 = \cos{\theta},
\end{align}
we retrieve the initialisation presented. This is convenient as we reduced the number of parameters to only $\theta$ to completely and uniquely define $\sigma_c$. Also it presents the option to work with another more natural parameter as in this case $f_1^2$ and $f_2^2$ are related by $f_1^2+f_2^2=1$. In fact, we just showed that the exact initialisation does not matter as long as $\sigma_c$ is uniquely defined and that we can choose the most convenient one. Note also that $f_2 = \sqrt{2} \cos{(\theta)}$ is also an option that keeps the norm of $\boldsymbol{a}$ independent of the DOC. 

\section{Complementary expressions}
\setcounter{table}{0}
\renewcommand{\thetable}{B\arabic{table}}

\label{sec:appendix_initialisations}
We display in Tab.~\ref{main_table} the main tools used to contract Eq.~(\ref{fulleqdMihdt}) with $\hat{r}_{ij}$. 

\begin{table*}
    \centering
    \begin{tabular}{c|c}
        Expression &  Reduced form \\ \hline \\ [-0.85em]
        $\hat{r}_{ii}$ & $1$ \\ [0.15em]
        $\hat{P}_{ii}$ & $2$ \\ [0.15em]
        $\hat{P}_{ij}\hat{r}_{il}$ & $0$ \\ [0.15em]
        $\hat{P}_{ij}\hat{P}_{il}$ & $\hat{P}_{jl}$ \\ [0.25em] \hline \\ [-0.85em]
        $\partial_i$ & $\hat{r}_i \partial_r$ \\ [0.15em]
        $\partial_i \partial_j$ & $\hat{r}_{ij} \partial_r^2 + \hat{P}_{ij} \frac{\partial_r}{r}$ \\ [0.15em]
        $\partial_i \hat{r}_j$ & $\frac{1}{r} \hat{P}_{ij}$ \\ [0.15em]
        $\partial_i \hat{r}_{jl}$ & $\frac{1}{r} (\hat{r}_l\hat{P}_{ij}+\hat{r}_j\hat{P}_{il}) $ \\ [0.15em]
        $\partial_i \hat{P}_{jl}$ & $-\frac{1}{r} (\hat{r}_l\hat{P}_{ij}+\hat{r}_j\hat{P}_{il}) $ \\ [0.25em] \hline \\ [-0.85em]
        $T_{ii}$ & $2T_\mathrm{N} + T_\mathrm{L}$ \\ [0.15em]
        $\hat{r}_i T_{ij}$ & $\hat{r_j} T_\mathrm{L}$ \\ [0.15em]
        $\hat{r}_{ij} T_{ij}$ & $T_\mathrm{L}$ \\ [0.15em]
        $\hat{P}_{ij} T_{ij}$ & $2T_\mathrm{N}$ \\ [0.15em]
        $M_{ij} T_{il}$ & $\hat{P}_{jl} M_\mathrm{N} T_\mathrm{N}+ \hat{r}_{jl} M_\mathrm{L} T_\mathrm{L}$ \\ [0.25em] \hline \\ [-0.85em]
        $T_{iijj}$ & $\overline{T}_\mathrm{L} + 8\overline{T}_\mathrm{N} + 4\overline{T}_\mathrm{LN}$ \\ [0.15em]
        $T_{iijl}$ & $\hat{r}_{jl}\overline{T}_\mathrm{L} + 4\hat{P}_{jl}\overline{T}_\mathrm{N} + (\hat{P}_{jl}+ 2\hat{r}_{jl})\overline{T}_\mathrm{LN}$ \\ [0.15em]
        $\hat{r}_i T_{ijhl}$ & $\hat{r}_{jhl}\overline{T}_\mathrm{L} + (\hat{P}_{jl}\hat{r}_{h}+\hat{P}_{jh}\hat{r}_{l}+\hat{P}_{hl}\hat{r}_{j})\overline{T}_\mathrm{LN}$ \\ [0.15em]
        $\hat{r}_{ij} T_{ijhl}$ & $\hat{r}_{hl}\overline{T}_\mathrm{L} + \hat{P}_{hl}\overline{T}_\mathrm{LN}$ \\ [0.15em]
        $\hat{P}_{ij} T_{ijhl}$ & $4\hat{P}_{hl}\overline{T}_\mathrm{N} + 2\hat{r}_{hl}\overline{T}_\mathrm{LN}$ \\ [0.25em] \hline \\ [-0.85em]
        $\partial_i T_{ijhl}$ & $\hat{r}_{jhl} (\partial_r \overline{T}_\mathrm{L} + \frac{2}{r} \overline{T}_\mathrm{L} -\frac{6}{r} \overline{T}_\mathrm{LN}) + (\hat{r}_j \hat{P}_{hl} + \hat{r}_h \hat{P}_{jl} + \hat{r}_l \hat{P}_{jh}) (\partial_r \overline{T}_\mathrm{LN} + \frac{4}{r} \overline{T}_\mathrm{LN} -\frac{4}{r} \overline{T}_\mathrm{N})$ \\ [0.5em]
        \multirow{2}{*}{$\partial_i \partial_j T_{ijhl}$} & $\hat{r}_{hl}(\partial_r^2 \overline{T}_\mathrm{L} + \frac{4}{r} \partial_r \overline{T}_\mathrm{L} -\frac{10}{r} \partial_r \overline{T}_\mathrm{LN} +\frac{2}{r^2}  \overline{T}_\mathrm{L}-\frac{22}{r^2}  \overline{T}_\mathrm{LN} + \frac{16}{r^2}\overline{T}_\mathrm{N})$  \\ [0.10em]
        &  $+\hat{P}_{hl}(\partial_r^2 \overline{T}_\mathrm{LN} +\frac{8}{r} \partial_r \overline{T}_\mathrm{LN}-\frac{4}{r}\partial_r \overline{T}_\mathrm{N} +\frac{12}{r^2}  \overline{T}_\mathrm{LN} - \frac{12}{r^2}\overline{T}_\mathrm{N})$  \\
    \end{tabular}
    \caption{Summary of the most basic tools to contract Eq.~(\ref{fulleqdMihdt}) with $\hat{r}_{ih}$.}
    \label{main_table}
\end{table*}

If we carry out all the algebra of Sec.~\ref{sec:secondini} we get the following expressions for the two-point velocity correlators
\begin{eqnarray}
\label{appendix:velocity_correlators}
&& \hspace{-0.4cm} T_{ij} = \frac{\tau b^2}{12} \bigg\{ \hat{r}_{ij} (\frac{\Omega + 1}{2} + \Omega \partial_z^2) + \hat{P}_{ij} (\frac{\Omega + 1}{2} + \Omega \frac{\partial_z}{z}) \bigg\} j_0(z), \nonumber \\
&& \hspace{-0.4cm}  T^{x^2y^2}_{ijhl} =\frac{\tau^2 b^4}{120} \bigg\{ \hat{r}_{ijhl} \bigl[ \frac{3}{16}\Omega_1 \partial_z^4 + \frac{3}{2}\Omega_2 \partial_z^2 + 3 \Omega_3 + 6\frac{\Omega_{\mathrm{tot}}}{j_0(2z)} \bigl] \nonumber \\
&& \hspace{0.1cm} + \hat{r}_{(ij}\hat{r}_{hl)} \bigl[ \frac{3}{16}\Omega_1 ( \frac{\partial_z^2}{z^2}-\frac{\partial_z}{z^3}) + \frac{1}{2}\Omega_2 \frac{\partial_z}{z} +  \Omega_3 +2\frac{\Omega_{\mathrm{tot}}}{j_0(2z)}\bigl] \nonumber \\
&& \hspace{0.1cm} + \hat{P}_{(ij}\hat{r}_{hl)} \bigl[ \frac{3}{16}\Omega_1 (\frac{\partial_z^3}{z} -2 \frac{\partial_z^2}{z^2}+2\frac{\partial_z}{z^3}) + \frac{1}{4}\Omega_2 (\partial_z^2+ \frac{\partial_z}{z}) \nonumber \\
&& \hspace{3cm}+  \Omega_3+2\frac{\Omega_{\mathrm{tot}}}{j_0(2z)} \bigl]\bigg\}  j_0(2z) , \nonumber \\
&& \hspace{-0.4cm} T^{x^3y}_{ijhl} = \frac{\tau^2 b^4}{40} \bigg\{ \hat{r}_{ijhl} \bigl[ 3\Omega_1 \partial_z^4 + 6\Omega_2 \partial_z^2 + 3 \Omega_3 \bigl] \nonumber \\
&& \hspace{0.1cm} + \hat{r}_{(ij}\hat{r}_{hl)} \bigl[ 3\Omega_1 ( \frac{\partial_z^2}{z^2}-\frac{\partial_z}{z^3}) + 2\Omega_2 \frac{\partial_z}{z} +  \Omega_3 \bigl] \nonumber \\
&& \hspace{0.1cm} + \hat{P}_{(ij}\hat{r}_{hl)} \bigl[ 3\Omega_1 (\frac{\partial_z^3}{z} -2 \frac{\partial_z^2}{z^2}+2\frac{\partial_z}{z^3}) + \Omega_2 (\partial_z^2+ \frac{\partial_z}{z}) \nonumber \\
&& \hspace{4cm} +  \Omega_3 \bigl]\bigg\}  j_0(z), 
\end{eqnarray}

Where the set of five parameters is defined as follow
\begin{eqnarray}
\label{appendix:param_set}
&& \hspace{-0.4cm} \begin{aligned}
    \Omega = 2\sin{(\theta)}^2 - 1, && \Omega_1 = \Omega^2, && \Omega_2 = \Omega \frac{\Omega+1}{2},
\end{aligned} \nonumber \\
    && \begin{aligned}
        \Omega_3 = \bigl(\frac{\Omega+1}{2}\bigl)^2,&& \Omega_{\mathrm{tot}}= \frac{\Omega_1}{5}-\frac{2\Omega_2}{3}+ \Omega_3.
    \end{aligned}
\end{eqnarray}

These five parameters appear very naturally in the derivation of the velocity correlators that is why we decided not to reduce the expressions to a single dependency on $\Omega$. Note that $\Omega=1$ for an incompressible flow and $\Omega=-1$ for a fully irrotational one.

\section{WKB solutions derivation}
\label{sec_appendix_wkb_sol}
The scaling solution that has been derived in the previous section only works if the term $\sqrt{\eta/\eta_t}$ is neglected. However to include effects due to a finite magnetic resistivity we should not systematically neglect it. A WKB approximation can be used to evaluate the solution of Eq.~(\ref{2ndcasesmallz}) including the finite resistivity.

\subsection{The WKB approximation}
The WKB (Wentzel–Kramers–Brillouin) approximation is first introduced in 1926 \citep{kramers1926wellenmechanik,wentzel1926verallgemeinerung}. In particular this approximation method has been extensively used in quantum mechanics to solve the Schrödinger equation \citep{merzbacher1961quantum,griffiths2018introduction}. Formally the method can be used to solve equations of the type 
\begin{equation}
\label{WKB_eq_gencons}
    \frac{\mathrm{d}^2 \Theta}{\mathrm{d}x^2}+p(x)\Theta =0,
\end{equation}
where the WKB solutions to this equation are linear combinations of 
\begin{equation}
    \Theta = \frac{1}{p^{1/4}} \exp \! \bigg( \pm i \int p^{1/2}~\mathrm{d}x \bigg) .
\end{equation}
We call turning points the value of $x$ where $p(x)$ is zero. In a given interval if $p(x)<0$ the solution is in the form of growing and decaying exponential however if $p(x)>0$ we have an oscillatory regime. Moreover the solutions need to satisfy boundary conditions, especially it is common to impose $\Theta(x)\rightarrow~ 0$ for $x \rightarrow \pm \infty$.

\subsection{Magnetic spectrum and growth rate at finite magnetic Reynolds number}

In the context of dynamos the WKB approximation is commonly used to derive the growth rate of the two-point correlation function of magnetic fluctuations. Reconsider Eq.~(\ref{2ndcasesmallz}), which is valid in the limit $z\ll 1$. To apply the WKB approximation we define a new coordinate, which is more convenient to use \citep{subramanian1997dynamics}, as $e^x = \Bar{z} \equiv \sqrt{\eta_t/\eta} z$. With this new coordinate Eq.~(\ref{2ndcasesmallz}) becomes 
\begin{eqnarray}
\label{WKBeq}
&&\hspace{-0.8cm}\bigg( \frac{\mathrm{d}^2 \tilde{M}_\mathrm{L}}{\mathrm{d}x^2} - \frac{\mathrm{d} \tilde{M}_\mathrm{L}}{\mathrm{d}x} \bigg) \bigg( \Bar{\tau}\epsilon(\theta) + \frac{5-\Omega}{5(\Omega+3)} + \frac{2}{\Bar{z}^2} \bigg) \nonumber\\
&&\hspace{0.5cm} +  \frac{\mathrm{d} \tilde{M}_\mathrm{L}}{\mathrm{d}x}  \bigg( 6\Bar{\tau}\epsilon(\theta) + 6\frac{5-\Omega}{5(\Omega+3)} + \frac{8}{\Bar{z}^2} \bigg) \nonumber\\
    && \hspace{1.3cm} + \tilde{M}_\mathrm{L}  \bigg( \Bar{\tau}\zeta(\theta) + \frac{8}{\Omega+3} -\gamma \bigg)=0.
\end{eqnarray}
To simplify notations we rewrite 
\begin{equation}
        \bigg( \frac{\mathrm{d}^2 \tilde{M}_\mathrm{L}}{\mathrm{d}x^2} - \frac{\mathrm{d} \tilde{M}_\mathrm{L}}{\mathrm{d}x} \bigg) A(x,\theta) +  \frac{\mathrm{d} \tilde{M}_\mathrm{L}}{\mathrm{d}x}  B(x,\theta) + \tilde{M}_\mathrm{L}  C(x,\theta) =0, 
\end{equation}
where the three functions are simply
\begin{eqnarray}
\label{eq_three_func}
A(x,\theta) &&\hspace{-0.3cm}= \Bar{\tau}\epsilon(\theta) + \frac{5-\Omega}{5(\Omega+3)} + \frac{2}{\Bar{z}^2},\nonumber\\
B(x,\theta) &&\hspace{-0.3cm}= 6\Bar{\tau}\epsilon(\theta) + 6\frac{5-\Omega}{5(\Omega+3)} + \frac{8}{\Bar{z}^2},\nonumber\\
C(x,\theta) &&\hspace{-0.3cm}= \Bar{\tau}\zeta(\theta) + \frac{8}{\Omega+3} -\gamma.
\end{eqnarray}
We further assume that $\tilde{M}_\mathrm{L}$ can be expressed as a product of two functions $\tilde{M}_\mathrm{L}=g(x)W(x)$. The idea is to impose certain relations on $g(x)$ such that all first order derivatives of $W(x)$ are cancelled, leading us to an equation that has the desired form. If we take 
\begin{equation}
\label{eqg}
    \frac{\mathrm{d} g}{\mathrm{d}x} = g \frac{A(x,\theta)-B(x,\theta)}{2A(x,\theta)},
\end{equation}
we find the desired equation Eq.~(\ref{WKB_eq_gencons}) for $W(x)$ with 
\begin{eqnarray}
\hspace{-0.3cm} p(x) = \frac{1}{A^2} \bigg[ AC- \frac{1}{2}(B'A-A'B)-\frac{1}{4}(A-B)^2 \bigg],
\end{eqnarray}
where primes denote derivative with respect to $x$ and the three functions are given by Eq.~(\ref{eq_three_func}). After some computation we can even show that 
\begin{equation}
\label{px}
    p(x) = \frac{A_0 \Bar{z}^4 - B_0  \Bar{z}^2-9}{(2+F  \Bar{z}^2)^2},
\end{equation}
where, for convenience, we set the following three functions of the DOC
\begin{eqnarray}
A_0 & =& \bigg( \Bar{\tau}\epsilon(\theta) + \frac{5-\Omega}{5(\Omega+3)} \bigg) \nonumber \\
&& \hspace{0.5cm} \times \biggl\{ \frac{7+5\Omega}{4(\Omega+3)}+ \Bar{\tau} \bigl[ \zeta(\theta)- \frac{25}{4}\epsilon(\theta)\bigl] -\gamma \biggl\} ,\nonumber \\
        B_0 &=& 2\gamma + 19 \Bar{\tau}\epsilon(\theta)-2\Bar{\tau} \zeta(\theta) + \frac{15-19\Omega}{5(\Omega+3)}, \nonumber\\
        F & =&  \Bar{\tau}\epsilon(\theta) + \frac{5-\Omega}{5(\Omega+3)} .
\end{eqnarray}
Recall that we are interested in the solution for the range $z_{\eta}~\ll~z~\ll~1$ which implies roughly that $1~\ll~\Bar{z}~\ll~R_\mathrm{M}^{1/2}$. If we take the limit of very small $\Bar{z}$, $x \rightarrow -\infty$, we see that $p\rightarrow -9/4$. As $\Bar{z}$ increases $p(x)$ increases too, let's call the first turning point $\Bar{z}_0$. We can guess from the evaluation of $\gamma$ in Sec.~\ref{sec:secondini} that $A_0$ is very small compared to $B_0$. Indeed when plugging in the value for $\gamma$ we found previously, we obtain that $A_0$ goes to zero while $B_0$ has a part independent on $\Bar{\tau}$. In particular it implies that $\Bar{z}_0$ is large enough to neglect the constant terms in the equation of $p(x)$ (i.e.~$\Bar{z}_0\gg 1$). The opposite limit of very large $\Bar{z}$, $x \rightarrow \infty$, is not described by Eq.~(\ref{WKBeq}) as it is valid only in the small $z$ limit. We need to go back to Eq.~(\ref{generalisedKazantsevequation}) and use that in the limit of very large $z$ the velocity correlators and their derivatives should go to zero. After some computation we obtain for the highest contribution
\begin{equation}
\label{px2}
    p(x) \sim -2e^{2x} \frac{(1+\eta_t/\eta) \gamma_0}{V(\theta,\eta_t,\eta,\Bar{\tau})^2},
\end{equation}
such that $p(x)<0$ in this limit. Note that we do not need to specify the exact form of $V(\theta,\eta_t,\eta,\Bar{\tau})$ as the denominator is always positive. In this formula we also neglected terms that depend on $\Bar{\tau}$ in the numerator as they should always be smaller than $\eta_t / \eta$ or $\gamma_0$ which are both positive.
Such a form means that $p(x)$ must have gone through another zero at some point that we call $\Bar{z}_1$. To simplify the treatment we will say that Eq.~(\ref{px}) is valid for $z<1$ and Eq.~(\ref{px2}) is valid for $z>1$. The boundary between the two can be taken to be $z_1$ such that $\Bar{z}_1 \sim R_\mathrm{M}^{1/2}$. In fact we will find that the final results have a small dependence on the exact value of $z_1$ such that we can approximate it without changing the conclusions \citep{schekochihin2002spectra,brandenburg2005astrophysical,bhat2015fluctuation}. To summarise we consider that we have damped solutions for $\Bar{z}\ll \Bar{z}_0$ and $\Bar{z}_1\ll \Bar{z}$ and an oscillatory one for $\Bar{z}_0\ll \Bar{z}\ll \Bar{z}_1$. The exponentially growing solutions are discarded as $M_\mathrm{L}(z)$ must remain finite at both $z=0$ and $z=\infty$.

In order for the oscillatory solution to match the two damped regime we have to require \citep{mestel1991galactic,jeffreys1999methods}
\begin{equation}
\label{cond}
    \int_{x_0}^{x_1} p(x)^{1/2}~\mathrm{d}x = \frac{(2n+1)\pi}{2},
\end{equation}
where $n$ is an integer. This condition is key to determine the growth rate $\gamma$ of the two-point correlation of the magnetic field. In the context of this work we only consider the fastest eigen-mode given by $n=0$. As we already mentioned the constant terms in Eq.~(\ref{px}) can be neglected which makes the solution to Eq.~(\ref{cond}) exact. Evaluating the integral gives
\begin{eqnarray}
\int_{x_0}^{x_1} p(x)^{1/2}~\mathrm{d}x &=& \int_{\Bar{z}_0}^{\Bar{z}_1} \frac{p(z)^{1/2}}{z}~\mathrm{d}z,\nonumber\\
        &&\hspace{-0.3cm}\simeq \int_{\Bar{z}_0}^{\Bar{z}_1} \frac{\sqrt{A_0 z^2- B_0}}{Fz^2}~\mathrm{d}z, \nonumber\\
        &&\hspace{-2.7cm}= \frac{\sqrt{A_0}}{F} \biggl\{ \ln{\bigg( \frac{z_1}{z_0} + \sqrt{\frac{z_1^2}{z_0^2}-1} \bigg)} -\sqrt{1-\frac{z_0^2}{z_1^2}} \biggl\},
\end{eqnarray}
where to go from first to second line we used that $\Bar{z}_0 \sim \sqrt{B_0/A_0 }>0$. We can thus use the condition of Eq.~(\ref{cond}), square both sides, and isolate the growth rate. The growth rate is finally given by 
\begin{eqnarray}
\label{gamma_final}
\gamma &=& -\bigg( \frac{\pi}{ \ln{(R_\mathrm{M})}}\bigg)^2 \bigg[ \frac{5-\Omega}{5(\Omega+3)} + \Bar{\tau}\epsilon(\theta) \bigg] \nonumber\\ 
&& \hspace{0.6cm} + \frac{7+5\Omega}{4(\Omega+3)} + \Bar{\tau} \bigg[ \zeta(\theta)- \frac{25}{4}\epsilon(\theta) \bigg], 
\end{eqnarray}
where again we plugged the self-consistent value for $\gamma_0$. Note that in this equation we also used the self-consistent evaluations $\Bar{z}_0 \sim \ln{(R_\mathrm{M})}$ and $\Bar{z}_1 \sim R_\mathrm{M}^{1/2}$, such that we neglected $\Bar{z}_0$ compared to $\Bar{z}_1$.

In the oscillatory range $1~\ll~\Bar{z}_0~\ll~\Bar{z}~\ll~\Bar{z}_1$ the WKB solution is thus given by
\begin{equation}
    W(x)\sim \bigg( \frac{\ln{(R_\mathrm{M})}}{ \pi}\bigg)^{1/2} \cos{\bigg[ \frac{\pi}{\ln{(R_\mathrm{M})}} \ln{\bigg( \frac{z}{z_0} \bigg)} \bigg]}.
\end{equation}
In this limit we see that Eq.~(\ref{eqg}) can be simplified such that $g'(x)\rightarrow -5g(x)/2$ which gives $g(x) \sim e^{-5x/2}$. The two-point magnetic correlation function is then also scaling as $z^{-5/2}$. So finally we find the equation for the longitudinal two-point magnetic correlation function in the region $z_{\eta}~\ll~z~\ll~1$
\begin{equation}
\label{MLwetestlplpp}
    M_\mathrm{L}(z,t) = e^{\gamma \tilde{t}} z^{-5/2} M_0 \cos{\bigg[ \frac{\pi}{\ln{(R_\mathrm{M})}} \ln{\bigg( \frac{z}{z_0} \bigg)} \bigg]},
\end{equation}
where $\gamma$ is given by Eq.~(\ref{gamma_final}).

\subsection{Validity of the WKB approximation}
\label{sec:validity}

\begin{figure*}
	\includegraphics[width=2\columnwidth]{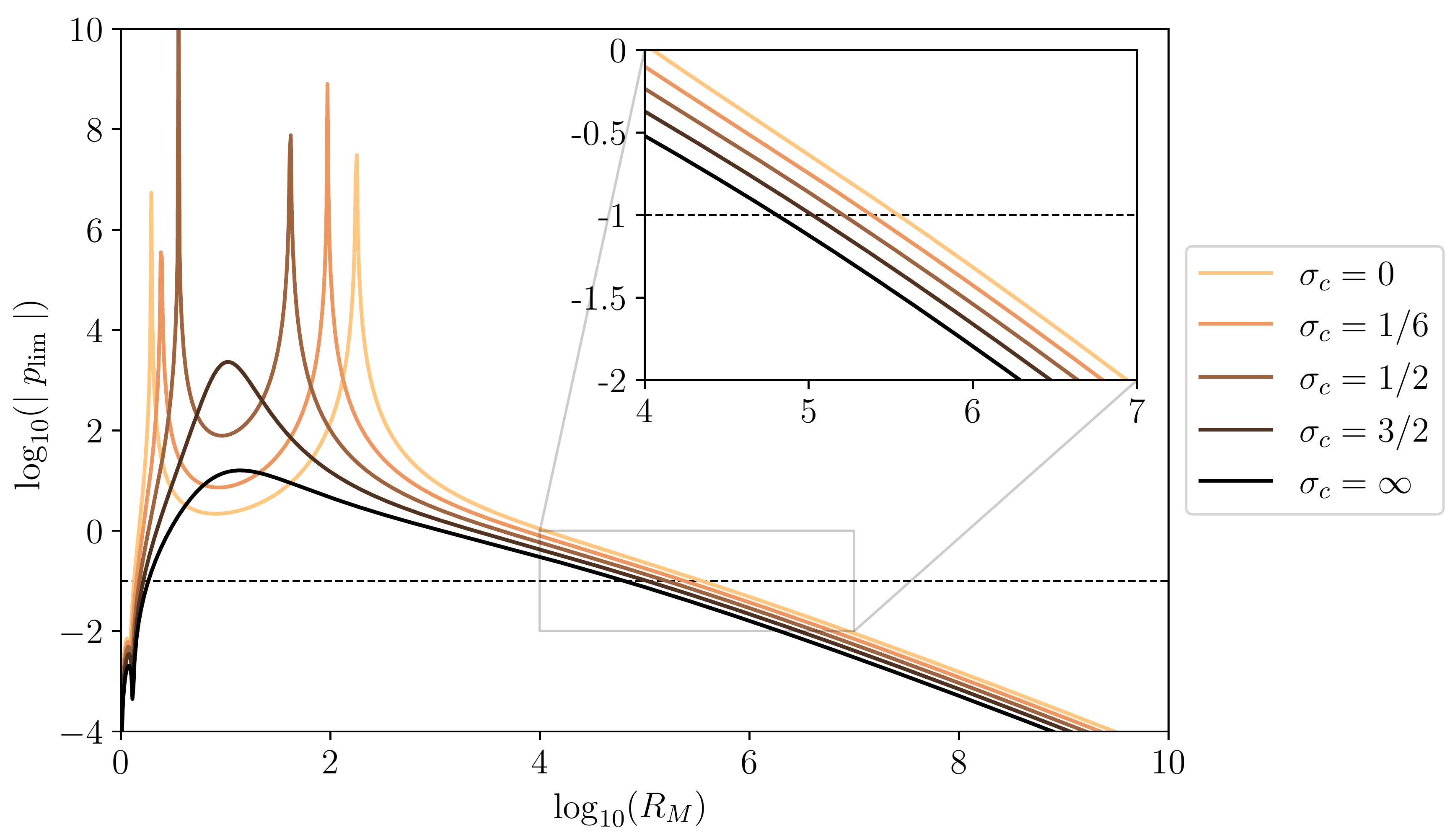}
    \caption{$p_{\mathrm{lim}}$ as a function of the magnetic Reynolds number $R_\mathrm{M}$ for $St=10^{-2}$ and a couple of DOC $\sigma_c$. The dashed line corresponds to the threshold over $p_{\mathrm{lim}}$ such that the intersections with the curves give the threshold $R_{\mathrm{M},\mathrm{thresh}}$ for which the WKB approximation holds.}
    \label{fig_RM_crit}
\end{figure*}

It can be showed that if we plug the solutions of the WKB approximation into Eq.~(\ref{WKB_eq_gencons}) we arrive at the following equation
\begin{equation}
    \frac{\mathrm{d}^2 \Theta}{\mathrm{d}x^2}+\bigg(1+ \frac{1}{4p(x)^2}\frac{\mathrm{d}^2 p}{\mathrm{d}x^2} -\frac{3}{16p(x)^3}\bigl(\frac{\mathrm{d} p}{\mathrm{d}x}\bigl)^2\bigg)p(x)\Theta =0,
\end{equation}
such that we retrieve the initial problem to solve only if 
\begin{eqnarray}
 p_\mathrm{lim} & \equiv& \frac{1}{4p(x)^2}\frac{\mathrm{d}^2 p}{\mathrm{d}x^2} -\frac{3}{16p(x)^3}\bigl(\frac{\mathrm{d} p}{\mathrm{d}x}\bigl)^2 \nonumber \\
 && \hspace{-0.3cm}= \frac{z^2 p''(z)+z p'(z)}{4p(z)^2} -\frac{3z^2 p'(z)^2}{16p(z)^3}
\end{eqnarray}
is very small compared to 1. Here primes denote derivatives with respect to the $z$ variable. Furthermore, in a similar way to \citet{schober2012magnetic}, we consider that the criterion of validity for our WKB approximation is $|p_\mathrm{lim}|<0.1$. We find that $p_\mathrm{lim}$ depends not only on the magnetic Reynolds number but also on $St$, $\sigma_c$ and $z_c$. We define here $z_c$ to be the scale at which we evaluate $p(z)$ and it derivatives. As the WKB approximation is valid between the two zeros of $p(z)$ we must impose $z_0 \ll z_c \ll z_1$. Until now, we only ask $R_\mathrm{M}$ to be very large, but the latter criterion gives us a way to quantify it. In particular, we use the expressions derived earlier for $p(z)$ and $\Bar{\tau}$ to define a threshold on $R_\mathrm{M}$ for which we consider that the derived results are valid\footnote{Note that this threshold on $R_\mathrm{M}$ is completely unrelated from the critical value of magnetic Reynolds number for which the dynamo can exist.}. In order to respect the conditions imposed on $z_c$, we take $z_c=(z_0+z_1)/2$. Although the scale can seem arbitrary, we find only a slight dependency on it as long as $z_c$ is not  too close to $z_0$ or $z_1$. In Fig.~\ref{fig_RM_crit} we present $p_\mathrm{lim}$ for a fixed $St=10^{-2}$ and a few DOC. Again, $St$ being tiny its exact value does not highly impact $R_{\mathrm{M},\mathrm{thresh}}$. It appears that the threshold of this work, regarding $R_\mathrm{M}$ is around $5\cdot 10^5$. More precisely, the $R_{\mathrm{M},\mathrm{thresh}}$ threshold decreases until it reaches $\sim 7\cdot 10^4$ when the DOC goes to infinity. The results derived in this work concerning the magnetic field are thus valid for most astrophysical objects where the fluctuation dynamo plays a major role. Note that from Fig.~\ref{fig_RM_crit} we also have a valid WKB approximation for very small $R_\mathrm{M}$. We can exclude this range of validity as we derived our generalised Kanzantsev equation (i.e. the expansion with respect to $St$) with the condition that $R_\mathrm{M}$ was a large number. 

\section{Proof of Eq.~(\ref{forMltoM})}
\label{appforMltoM}

Let's start by expressing the magnetic power spectrum as the Fourier transform of the magnetic two-point correlation and take the Fourier transform of this expression

\begin{equation}
    M_k(k)= 2\pi k^2 \hat{M}_{ii}(k) = \frac{k^2}{(2\pi)^2}\int M_{ii}(r) e^{i\boldsymbol{k}\cdot \boldsymbol{r}}~\mathrm{d}^3\boldsymbol{r}.
\end{equation}
Now use the properties of $M(r)$ to derive the following
\begin{eqnarray}
M_k(k) &=& \frac{k^2}{2\pi}\int r^2 \sin{(\theta)} M_{ii}(r) e^{ikr \cos{(\theta)}}~\mathrm{d}r\mathrm{d}\theta, \nonumber \\
    && \hspace{-0.3cm}= \frac{ik}{2\pi}\int r M_{ii}(r) \bigg( e^{-ikr}- e^{ikr} \bigg)~\mathrm{d}r, \nonumber \\
    && \hspace{-0.3cm}= \frac{1}{\pi}\int kr \bigg( 3M_\mathrm{L}(r) + r\partial_r M_\mathrm{L}(r) \bigg) \sin{(kr)}~\mathrm{d}r, \nonumber \\
    && \hspace{-0.3cm}= \frac{1}{\pi}\int kr M_\mathrm{L}(r)  \bigg( \sin{(kr)}- \cos{(kr)} \biggl)~\mathrm{d}r, \nonumber \\
    && \hspace{-0.3cm}= \frac{1}{\pi}\int (kr)^3 M_\mathrm{L}(r)  j_1(kr)~\mathrm{d}r, 
\end{eqnarray}
where to go from the third to fourth line we integrated by parts. It is pretty obvious from the definition of $rM_\mathrm{L}(r)$ that the boundary terms just go to zero.

\bibliography{apssamp}

\end{document}